\documentclass[a4paper,useAMS,usenatbib,usegraphicx]{mn2e} 
\usepackage{amsfonts,amsmath,amssymb}
\usepackage{times}
\usepackage{slashbox}

\newcommand{\changed}[1]{{#1}}

\newcommand{\lx}{\hbox{$L_\mathrm{X}$}}
\newcommand{\lxt}{\hbox{$L_\mathrm{X}$-$T$}}
\newcommand{\lxtx}{\hbox{$L_\mathrm{X}$-$T_{\rm X}$}}
\newcommand{\lxtsl}{\hbox{$L_\mathrm{X}$-$T_{\rm sl}$}}

\newcommand{\knormtsl}{\hbox{$K(r_{1000})$-$T_{\rm sl}$}}
\newcommand{\kshapetsl}{\hbox{$K(r_{1000})/K(0.1r_{200})$-$T_{\rm sl}$}}

\def\Msun{\hbox{$\rm\, M_{\odot}$}}

\newcommand{\lgalaxies}{L-Galaxies}
\newcommand{\SAM}{SA model}
\newcommand{\spose}[1]{{\hbox to 0pt{#1\hss}}}
\newcommand{\lta}{\mathrel{\spose{\lower 3pt\hbox{$\mathchar"218$}}
     \raise 2.0pt\hbox{$\mathchar"13C$}}}

\begin{document}

\title[Heating and enriching the intracluster medium]{Heating and
  enriching the intracluster medium} 
\author[C.~J.~Short et al.]{C.~J.~Short$^1$, P.~A.~Thomas$^1$\thanks{E-mail: P.A.Thomas@sussex.ac.uk} and O.~E.~Young$^1$\\ 
$^1$Astronomy Centre, University of Sussex, Falmer, Brighton, BN1 9QH, United Kingdom}
\maketitle

\begin{abstract}
We present numerical simulations of galaxy clusters with stochastic
  heating from active galactic nuclei (AGN) that are able to reproduce the
  observed entropy and temperature profiles of non-cool-core (NCC) clusters.
Our study uses $N$-body hydrodynamical simulations to investigate how star
formation, metal production, black hole accretion, and the associated feedback
from supernovae and AGN, heat and enrich diffuse gas in galaxy clusters. We
assess how different implementations of these processes affect the thermal and
chemical properties of the intracluster medium (ICM), using high-quality X-ray
observations of local clusters to constrain our models. For the purposes of this
study we have resimulated a sample of $25$ massive galaxy clusters extracted
from the Millennium Simulation. Sub-grid physics is handled using a
semi-analytic model of galaxy formation, thus guaranteeing that the source of
feedback in our simulations is a population of galaxies with realistic
properties. We find that supernova feedback has no effect on the entropy and
metallicity structure of the ICM, regardless of the method used to inject energy
and metals into the diffuse gas. By including AGN feedback, we are able to
explain the observed entropy and metallicity profiles of clusters, as well as
the X-ray luminosity-temperature scaling relation for NCC systems. A
stochastic model of AGN energy injection motivated by anisotropic jet
heating -- presented for the first time here -- is crucial for this success.

With the addition of metal-dependent radiative cooling, our model is also able
to produce CC clusters, without over-cooling of gas in dense, central regions.
\end{abstract}


\section{Introduction}

\subsection{Background}

Clusters of galaxies are believed to be the largest gravitationally-bound
objects in the Universe. Their deep gravitational potential well means that the
largest clusters are 'closed boxes', in the sense that baryons ejected from
cluster galaxies by supernova (SN) explosions and active galactic nuclei (AGN)
do not escape the cluster completely, but instead end up in the hot, diffuse
plasma that fills the space between cluster galaxies -- the intracluster medium
(ICM).

The thermal properties of intracluster gas, which can be measured with X-ray
telescopes such as \emph{Chandra}, \emph{XMM-Newton} and \emph{Suzaku}, thus
provide a unique fossil record of the physical processes important in galaxy and
galaxy cluster formation and evolution, such as radiative cooling, star
formation, black hole accretion and the subsequent feedback from supernovae
(SNe) and AGN. In addition, measurements of the ICM chemical abundances yield
information about the production of heavy elements in stars in member galaxies,
providing constraints on nucleosynthesis, and the processes responsible for
their transport into the ICM.

A key diagnostic of the thermal state of intracluster gas is provided by the gas
entropy\footnote{We define the gas entropy as $K=k_{\rm
    B}T/n_\mathrm{e}^{\gamma-1}$, where $k_{\rm B}$ is Boltzmann's constant, $T$
  is the gas temperature, $n_\mathrm{e}$ is the electron number density and
  $\gamma=5/3$ is the ratio of specific heats for a monoatomic ideal
  gas.}. Entropy remains unchanged under adiabatic processes, such as
gravitational compression, but increases when heat energy is introduced and
decreases when radiative cooling carries heat energy away, thus providing an
indicator of the non-gravitational processes important in cluster formation.

In recent years, spatially-resolved observations have facilitated a detailed
examination of the radial distribution of entropy in clusters. Observed entropy
profiles are typically found to scale as $K\propto r^{1.1-1.2}$ at large
cluster-centric radii, $r\gtrsim 0.1r_{200}$\footnote{We define $r_{\Delta}$ as
  the radius of a spherical volume within which the mean matter density is
  $\Delta$ times the critical density at the redshift of interest. The mass
  enclosed within this sphere is denoted by $M_{\Delta}$.}
(e.g. \citealt{PSF03,SVD09,CDV09,SOP09,PAP10}). This power-law scaling agrees
with that predicted by simple analytical models based on spherical collapse
\citep{TON01} and cosmological hydrodynamical simulations that include
gravitational shock heating only (e.g. \citealt[hereafter STY10]{VKB05,
  NKV07,STY10}).

However, in the inner regions of clusters, observations have unveiled the
presence of a mass-dependent entropy excess with respect to theoretical
expectations, and a large dispersion in central entropy values
(e.g. \citealt{PCN99,LPC00,PSF03,PAP06,MOE07,CDV09,PAP10}). The source of this
entropy excess is likely to be a combination of non-gravitational heating from
astrophysical sources, such as SNe and AGN, and cooling processes. There is
evidence that the distribution of central entropy values is bimodal, with
morphologically disturbed non-cool-core (NCC) systems having an elevated central
entropy compared to dynamically relaxed cool-core (CC) systems
\citep[e.g.~][]{CDV09}. It is thought that the association of unrelaxed
morphology with a high central entropy is an indication that either cool cores
are destroyed by mergers, or that cool cores have never been able to form in
these objects.

The chemical state of the ICM is characterised by its metallicity, the
proportion of chemical elements present heavier than H and He. X-ray
spectroscopy of galaxy clusters has revealed emission features from a variety of
chemical elements, O, Ne, Mg, Si, S, Ar, Ca, Fe and Ni, all of which are
synthesised in stars and transported to the ICM by processes such as
ram-pressure stripping, galactic winds and AGN outflows. Type Ia SNe produce a
large amount of Fe, Ni, Si, S, Ar and Ca, but compared to Type II SNe, they only
produce very small amounts of O, Ne and Mg. Type Ia SN products are found to
dominate in cluster cores, whereas Type II SN products are more evenly
distributed (\citealt{FDP00,TBK01}; see \citealt{BOW10} for a recent
review). This can be explained by early homogeneous enrichment by Type II SNe,
which produce $\alpha$-elements in the protocluster phase, and a subsequent,
more centrally-peaked enrichment by Type Ia SNe, which have longer delay times
and continue to explode in the central cD galaxy long after the cluster is
formed.

The first detailed measurements of spatial abundance distributions were made by
\citet{DEM01}, using data from \emph{BeppoSAX}, who measured the radial Fe
abundance profiles for a sample of massive clusters. They found that CC clusters
have a sharp Fe abundance peak in central regions, whereas NCC clusters have
flat Fe abundance profiles. Subsequent observations with \emph{XMM-Newton} and
\emph{Chandra} have confirmed this dichotomy between the metallicity
distribution in CC and NCC clusters
\citep{TKH04,VMM05,PBC07,BEM07,MJF08,LEM08,MAT11}.


It is thought that cluster mergers, and the subsequent mixing of intracluster
gas, are responsible for destroying the central abundance peak found in CC
clusters. However, some systems with a highly disturbed morphology are also
found to have a high central metallicity. \citet[see also
\citealt{ROM10}]{LRM10} suggest that these objects correspond to relaxed CC systems
that have undergone a major merger, or a significant AGN heating event, very
recently, so that mixing processes have not yet had sufficient time to fully
erase low-entropy gas and the central abundance peak.

Explaining the observed thermal and chemical properties of the ICM from a
theoretical perspective requires a detailed understanding of the complex
interplay between large-scale gravitational dynamics and the various small-scale
astrophysical processes mentioned above. Numerical cosmological hydrodynamical
simulations have emerged as the primary tool with which to tackle this
problem. There has been considerable effort to include the processes relevant
for cluster formation and evolution in simulations in a self-consistent manner;
see \citet{BOK09} for a recent review. However, an explicit treatment is
unfeasible since these processes all occur on scales much smaller than can be
resolved with present computational resources.

Hydrodynamical simulations that include models of radiative cooling, star
formation, metal production and galactic winds generally fail to reproduce
observed ICM temperature, entropy and metallicity profiles. Simulated
temperature profiles typically have a sharp spike at small cluster-centric
radii, followed by a rapid drop in temperature moving further into the core
(e.g. \citealt{VAL03,TBS03,BMS04,RSP06,SSD07,NKV07}), in clear conflict with the
smoothly declining (flat) profiles of observed CC (NCC) clusters
(e.g. \citealt{SPO06,VKF06,APP09}). This is due to the adiabatic compression of
gas flowing in from cluster outskirts to maintain pressure support, following
too much gas cooling out of the hot phase. This over-cooling causes excessive
star formation in cluster cores, with predicted stellar fractions being about a
factor of $2$ larger than the observed value of $\sim 10\%$
\citep{BPB01,LMS03,BMBE08}, which in turn leads to excessive Fe production in
central regions, generating steeper abundance profiles than observed
(e.g. \citealt{VAL03,TBM04,RSP06,TBD07,DOS08}).

It is generally accepted that the solution to the over-cooling problem in
hydrodynamical simulations is extra heat input from AGN. Simple analytical
arguments convincingly show that the energy liberated by accretion onto a
central super-massive black hole is sufficient to suppress gas cooling and thus
quench star formation. The precise details of how this energy is transferred to
the ICM are not well understood at present, but it appears that there are two
major channels via which black holes interact with their surroundings (see
\citealt{MCN07} for a review).

At high redshift, mergers of gas-rich galaxies occur frequently and are expected
to funnel copious amounts of cold gas towards galactic centres, leading to high
black hole accretion rates and radiating enough energy to support the
luminosities of powerful quasars. Quasar-induced outflows have been
observationally confirmed in a number of cases
(e.g. \citealt{CBG03,CKG03,PRK03,GAB08,DBA10}).

Evidence for another mode of AGN feedback, not related to quasar activity, can
be seen in nearby CC clusters, which often contain radio-loud X-ray cavities in
the ICM. These bubbles are thought to be inflated by relativistic jets launched
from the central super-massive black hole
\citep{BSM01,BRM04,MNW05,FST06,MIY06,JHP08,GBT09,DRM10,GOV11}. Bubbles may rise
buoyantly, removing some of the central cool, enriched gas and allowing it to
mix with hotter gas in the outer regions of groups and clusters. Together with
the accompanying mechanical heating, this can constitute an efficient mechanism
for suppressing cooling flows, and redistributing metals throughout the
ICM. Such flows are seen in simulations of idealised clusters, performed with
hydrodynamical mesh codes
(e.g.~\citealt{CBK01,QBB01,RUB02,BKC02,BRU03,DBT04,RBR07,BRS09}).

Various authors have implemented self-consistent models of black hole growth and
AGN feedback in cosmological simulations of galaxy groups and clusters (in
addition to cooling, star formation, and thermal and chemical feedback from
SNe). \citet{SDH05a} developed a model for quasar mode AGN feedback (see also
\citealt{DSH05}), which was used in cosmological simulations of $10$ galaxy
groups by \citet{BDK08}. A model for radio mode AGN feedback based on bubble
injection was proposed by \citet{SIS06}, which was subsequently extended by
\citet{SSD07} to include quasar mode AGN feedback as well. Both \citet{SIS06}
and \citet{SSD07} performed cosmological simulations of a few massive clusters
with their respective models.

These studies demonstrated, in a qualitative manner, that AGN feedback is
effective in reducing the amount of cold baryons and star formation in the
central regions of groups and clusters. Furthermore, the gas density is reduced
and the temperature is increased, elevating the central entropy. \citet{SSD07}
also showed that AGN outflows drive metals from dense, star-forming regions to
large radii, flattening ICM abundance profiles relative to those predicted by a
run without AGN feedback. Such trends are precisely what is required to
reconcile simulations of galaxy clusters with observations.

A more quantitative assessment of the impact of AGN feedback on the ICM was
conducted by \citet{PSS08}. They resimulated a sample of $21$ groups and
clusters with the scheme of \citet{SSD07}, finding that the model could
reproduce the observed X-ray luminosity-temperature scaling relation, at least
on average. However, since their sample size is quite small, it is unclear
whether the model can generate a realistic population of CC and NCC systems and
thus explain the observed scatter about the mean relation. In addition, the
stellar fraction within the virial radii of their simulated objects appears
larger than \changed{observed.}

Another detailed study was undertaken by \citet{FBT10}, who resimulated a sample
of groups and clusters in a cosmological setting, using a model closely related
to that of \citet{SSD07}, but with a different implementation of radio mode AGN
feedback.  On group scales, they found that AGN heating was able to successfully
balance radiative cooling, reproducing observed stellar fractions, but the
central entropy (at $r_{2500}$) was about a factor of $2$ too high. In addition,
their predicted group Fe abundance profiles are flat for $r\gtrsim 0.3r_{500}$,
whereas observed profiles have a negative gradient out to the largest radii for
which measurements are possible (e.g. \citealt{RAP09}). There is also an
indication that the Fe distribution may be too sharply peaked in central regions
compared to observations. The effect of AGN feedback on galaxy groups was also
investigated by \citet{MSP10}, who implemented the AGN feedback scheme of
\citet{BOS09} in a cosmological simulation. With this model they were able to
explain the observed entropy, temperature and Fe abundance profiles of groups,
as well as observed X-ray scaling relations.

For massive clusters, \citet{FBT10} showed that their model can reproduce the
entropy structure of the ICM, but a factor of $3$--$4$ too many stars were
formed. The cluster Fe abundance profiles they obtained have a shape consistent
with that of observed profiles, although with a higher normalisation, but the
central Fe abundance may be over-estimated. \citet{DDT11} also examined the role
of AGN feedback in establishing the properties of the ICM, using a cosmological
AMR simulation of a massive cluster with a prescription for jet heating by
AGN. The entropy profile of their cluster agrees well with that of observed CC
clusters if metal-cooling is neglected, and when metals are allowed to
contribute to the radiative cooling, the resulting profile resembles that of a
NCC cluster instead. However, the metallicity profile of their cluster appears
steeper than observed.

\subsection{This work}

In this work, we pursue a different, but complementary, approach to the
theoretical study of galaxy clusters. Instead of undertaking self-consistent
hydrodynamical simulations, we adopt the hybrid approach of \citet[hereafter
  SHT09]{SHT09} which couples a semi-analytic model (\SAM) of galaxy formation
to a cosmological $N$-body/smoothed-particle hydrodynamics (SPH) simulation. In
this model, the energy imparted to the ICM by SNe and AGN is computed from a
\SAM\ and injected into the baryonic component of a non-radiative hydrodynamical
simulation; see SHT09 for details. The main advantage of this approach is that
feedback is guaranteed to originate from a realistic population of galaxies,
since \SAM{}s are tuned to reproduce the properties of observed galaxies. As a
consequence, the stellar fraction in massive clusters agrees with observations
\citep{YTS11}, which is not the case in self-consistent hydrodynamical
simulations.

We have extended the model of SHT09 to follow the metal enrichment of the ICM.
Note that \citet[see also \citealt{CTT08}]{COR06} have already used a similar
hybrid technique to study the pollution of intracluster gas by heavy
elements. However, they did not include energy injection from SNe and AGN, which
are likely to affect the distribution of metals in the ICM.

In the model of SHT09, the energy liberated by SN explosions and black
hole accretion is assumed to be distributed uniformly throughout the
diffuse gas of the host halo.  With this rather \emph{ad hoc} heating
model they were able to reproduce observed X-ray scaling relations for
NCC clusters, but ICM entropy profiles were found to be flatter than
observed within 0.5 times $r_{500}$ (STY10).  These simulations do not
well resolve the core ($r \lta 0.1\,r_{500}$), nor do they include
radiative cooling that is likely to be important in this region, at
least for CC clusters.  However, we would expect that they should be
able to provide a much better fit to X-ray observations of NCC
clusters outside the core.

The primary goal of this paper is, therefore, to formulate a new
feedback model that has a clear physical motivation and that is better
able to explain the radial variaton of both the thermal and chemical
properties of intracluster gas outside the core of the cluster.  To
help us do this we test a wide variety of different models for SN and
AGN feedback and metal enrichment, using a selection of X-ray data
(namely, entropy and metallicity profiles and the
luminosity-temperature scaling relation) to identify the features that
a model should possess in order to reproduce the data.

Our conclusion is that a stochastic heating model, motivated by observations of
anisotropic AGN outflows, provides a better fit to the observed properties of
the ICM than more commonplace models, such as heating a fixed number of
neighbours or heating particles by a fixed temperature. Using entirely plausible
duty cycles and opening angles for the jets, it is possible to provide an
acceptable fit to all available observations with our model.

Note that the use of SA models means that the feedback is not directly
coupled to the cooling of the gas -- that is why our previous work and
the bulk of this paper uses non-radiative simulations and restricts
its attention to NCC clusters.  However, towards the end of the paper
we introduce radiative cooling in an attempt to reproduced CC
clusters.  We estimate the degree to which the SA model fails to
supply the required feedback energy and show that there can be a
substantial short-fall at high redshift, but that it averages to under
10 per cent over the lifetime of the cluster.  We are able to
qualitatively reproduce some CC profiles, but we do not provide a
detailed quantitative analysis here.

In this work, we neglect many physical effects such as magnetic
fields, cosmic rays, thermal conduction, turbulent mixing, etc..  Our
principal reason for doing this is to keep the model simple and ease
interpretation of our results.  Some of these may be important in the
central regions of CC clusters ($r\lta r_{500}$) but there is little
evidence that they play a significant role at the larger radii that we
use to constrain our models.  We discuss this further at the end of
the paper.

The layout of this paper is as follows. In Section \ref{sec:sims}, we
present the details of our hybrid numerical model and describe our
cluster simulations. We investigate the effect of SN feedback on the
thermal and chemical properties of the ICM in Section \ref{sec:sne},
and assess how our results are affected by different choices of SN
feedback and metal enrichment models. We show, in agreement with
previous work, that SN have little impact on the entropy structure of
the intracluster gas.  In Section \ref{sec:agn} we examine the impact
of additional heating from AGN: these can reproduce the correct
scaling relations but give entropy profiles that are too flat.  Our
results motivate a new, stochastic feedback model based on jet
heating, which is described in Section \ref{sec:newagn}. In this
section, we also discuss what this model predicts for the thermal and
chemical properties of the ICM, and we conduct an exhaustive
comparison with observational data in Section \ref{sec:obscomp}. In
Section \ref{sec:cooling} we demonstrate that our model is capable of
producing both CC and NCC clusters with the inclusion of
metal-dependent radiative cooling. Our conclusions are presented in
Section \ref{sec:conc}.

For those readers who are mostly interested in the final model itself, rather
than the steps used to motivate it, we recommend skipping Sections~\ref{sec:sne}
and \ref{sec:agn}, at least on first reading.

\section{Simulations}
\label{sec:sims}

We make use of hydrodynamical resimulations of a sample of massive galaxy
clusters extracted from the dark-matter-only Millennium Simulation
\citep{SWJ05}. Our sample consists of $25$ objects with $9\times
10^{13}h^{-1}\Msun\lesssim M_{500}\lesssim 7\times 10^{14}h^{-1}\Msun$ and forms
a subset of the larger sample of $337$ groups and clusters resimulated by STY10
for their so-called FO simulation, one of the Millennium Gas
Simulations\footnote{The Millennium Gas Simulations are a series of
  hydrodynamical simulations designed to add gas to the dark matter structures
  found in the Millennium Simulation \citep{SWJ05}. At present, there are $3$
  simulations, each of which employs a different physical mechanism for raising
  the entropy of intracluster gas. The first of these is a reference model that
  includes gravitational heating only (the GO run). The second includes
  radiative cooling and uniform preheating at $z=4$ as a simple model for
  heating from astrophysical sources (the PC run). The third simulation is the
  FO run, where feedback from galaxies is computed from a \SAM\ using the hybrid
  model of SHT09.}. See STY10 for details of the cluster selection
procedure. Basic properties of our clusters are listed in Table
\ref{tab:clusprop}.

\begin{table}
\caption{The masses, $M$ (in units of $h^{-1}\Msun$), and dynamical
  temperatures, $k_{\rm B}T_{\rm dyn}$ (in units of keV), of the $25$ clusters
  used in this study within $r_{500}$ (second and third columns, respectively),
  and $r_{200}$ (third and fourth columns, respectively). Cluster C1 is our
  fiducial cluster, used for most of the plots in this paper.}
\label{tab:clusprop}
\begin{tabular}{@{}lcccc}
\hline
Cluster name & $M_{500}$ & $T_{{\rm dyn},500}$ & $M_{200}$ & $T_{{\rm dyn},200}$\\
\hline
C1 & $2.7\times 10^{14}$ & $3.9$ & $4.6\times 10^{14}$ & $3.9$ \\
\hline
C2 & $7.1\times 10^{14}$ & $7.5$ & $1.1\times 10^{15}$ & $7.1$ \\
C3 & $4.2\times 10^{14}$ & $5.8$ & $5.8\times 10^{14}$ & $5.4$ \\
C4 & $3.5\times 10^{14}$ & $5.2$ & $4.9\times 10^{14}$ & $4.8$ \\
C5 & $3.9\times 10^{14}$ & $5.6$ & $6.4\times 10^{14}$ & $5.3$ \\
C6 & $7.3\times 10^{14}$ & $9.6$ & $1.1\times 10^{15}$ & $8.6$ \\
C7 & $5.7\times 10^{14}$ & $7.0$ & $9.0\times 10^{14}$ & $6.5$ \\
C8 & $5.0\times 10^{14}$ & $5.7$ & $7.1\times 10^{14}$ & $5.3$ \\
C9 & $3.7\times 10^{14}$ & $5.0$ & $5.0\times 10^{14}$ & $4.7$ \\
C10 & $3.9\times 10^{14}$ & $5.2$ & $5.2\times 10^{14}$ & $4.8$ \\
C11 & $2.8\times 10^{14}$ & $4.3$ & $4.1\times 10^{14}$ & $4.0$ \\
C12 & $3.4\times 10^{14}$ & $4.7$ & $5.2\times 10^{14}$ & $4.5$ \\
C13 & $3.5\times 10^{14}$ & $4.7$ & $5.1\times 10^{14}$ & $4.5$ \\
C14 & $3.6\times 10^{14}$ & $4.9$ & $5.9\times 10^{14}$ & $4.6$ \\
C15 & $2.6\times 10^{14}$ & $4.3$ & $4.2\times 10^{14}$ & $3.9$ \\
C16 & $3.3\times 10^{14}$ & $4.6$ & $4.5\times 10^{14}$ & $4.3$ \\
C17 & $2.4\times 10^{14}$ & $4.1$ & $4.1\times 10^{14}$ & $4.0$ \\
C18 & $2.3\times 10^{14}$ & $3.9$ & $3.7\times 10^{14}$ & $3.5$ \\
C19 & $2.2\times 10^{14}$ & $3.5$ & $3.2\times 10^{14}$ & $3.2$ \\
C20 & $1.7\times 10^{14}$ & $3.6$ & $3.5\times 10^{14}$ & $3.4$ \\
C21 & $1.7\times 10^{14}$ & $3.1$ & $2.3\times 10^{14}$ & $2.8$ \\
C22 & $1.6\times 10^{14}$ & $2.9$ & $2.2\times 10^{14}$ & $2.6$ \\
C23 & $9.8\times 10^{13}$ & $2.5$ & $1.9\times 10^{14}$ & $2.3$ \\
C24 & $1.1\times 10^{14}$ & $2.2$ & $1.6\times 10^{14}$ & $2.1$ \\
C25 & $8.7\times 10^{13}$ & $1.9$ & $1.3\times 10^{14}$ & $1.8$ \\
\hline
\end{tabular}

\medskip
We define dynamical temperature as $T_\mathrm{dyn}=\mu m_\mathrm{H}\langle
v^2\rangle/3k_\mathrm{B}$, where $\mu m_\mathrm{H}\approx10^{-27}$\,kg is the
mean particle mass and $\langle v^2\rangle$ is the mean square velocity.
\end{table}

Following STY10, the feedback model we adopt in our simulations is the hybrid
scheme of SHT09, where a \SAM\ of galaxy formation is used to compute the number
of stars formed and the energy transferred to the ICM by SNe and AGN. We refer
the reader to STY10 for a full description of the modelling process and
simulation parameters.

Briefly, we first perform dark-matter-only simulations of each region containing
a cluster in our sample using the massively parallel TreePM $N$-body/SPH code
GADGET-2 \citep{SPR05}. Virialised dark matter haloes are identified at each
simulation output using the friends-of-friends (FOF) algorithm, with a standard
linking length of $20\%$ of the mean inter-particle separation
\citep{DEF85}. Only groups with at least $20$ particles are kept, yielding a
minimum halo mass of $1.7\times 10^{10}h^{-1}\Msun$. Gravitationally bound
substructures orbiting within these FOF haloes are then found with a parallel
version of the SUBFIND algorithm \citep{SWT01}. From the stored subhalo
catalogues we construct dark matter halo merger trees by exploiting the fact
that each halo will have a unique descendant in a hierarchical scenario of
structure formation; see \citet{SWJ05} for further details.

The second stage is to generate galaxy catalogues for each resimulated region by
applying the Munich \lgalaxies\  \SAM\ of \citet{DLB07} to the halo merger trees. A
full description of the physical processes incorporated in \lgalaxies\ and model
parameters is given in \citet{CSW06} and \citet{DLB07}. For each galaxy in these
catalogues, we use its merger tree to compute the change in stellar mass,
$\Delta M_*$, and mass accreted by the central black hole, $\Delta M_{\rm BH}$,
between successive model outputs.  Knowledge of $\Delta M_*$ enables us to
incorporate star formation in our simulations as described below. From $\Delta
M_*$ and $\Delta M_{\rm BH}$ we can also calculate the energy imparted to
intracluster gas by Type II SNe, $\Delta E_{\rm SN}$, and AGN, $\Delta E_{\rm
  AGN}$, respectively. Details are given in SHT09.

For the purpose of this work, we have extended the model of SHT09 to also follow
the enrichment of the ICM by metals ejected from galaxies in winds. In the
\lgalaxies\ \SAM\, $43\%$ of the mass of newly formed stars is instantaneously
returned, and deposited in the cold gas disc of the host galaxy
\citep{CSW06}. In other words, the model assumes that metal ejection is
instantaneous and does not distinguish between emission from Type II and prompt
Type Ia SNe, and that from delayed Type Ia SNe and AGB stars. This will be added
in future work.

In each model galaxy, metals can reside in several distinct phases: stars, cold
disc gas, hot halo gas and gas ejected by winds from the halo into an external
`reservoir'. Only the latter two are relevant for the ICM. We define the total
mass in metals in diffuse gas to be
\begin{equation}
M_{\rm Z,ICM}=M_{\rm Z,hot}+M_{\rm Z,ej},
\end{equation}
where $M_{\rm Z,hot}$ and $M_{\rm Z,ej}$ are the mass in metals in hot and
ejected gas, respectively.

Once a galaxy falls into a FOF group, becoming a satellite of the central galaxy
of the halo, all of its metals in hot and ejected gas are assumed to be
associated with the central galaxy. It follows that $M_{\rm Z,ICM}$ is non-zero
only for central galaxies. Given a particular halo at some output redshift $z_n$, we
compute the change in metal content of the ICM, $\Delta M_{\rm Z,ICM}$, since
the previous output, $z_{n+1}$, by taking $M_{\rm Z,ICM}$ for the central galaxy
and subtracting the sum of $M_{\rm Z,ICM}$ for every galaxy that is a progenitor
of any galaxy contained in the host FOF group and also a central galaxy of a
halo at $z_{n+1}$:
\begin{equation}
\Delta M_{\rm Z,ICM}=M_{\rm Z,ICM}(z_n)-\sum_\mathrm{prog.}M_{\rm Z,ICM}(z_{n+1}).
\end{equation} 
The quantity $\Delta M_{\rm Z,ICM}$ is used to implement metal enrichment of the
ICM in our simulations, as described in subsequent sections.

Finally, we couple the \lgalaxies\ \SAM\ to hydrodynamical simulations
of our clusters to track the effect of feedback from galaxies on the
thermal and chemical properties of the ICM. The initial conditions for
these resimulations are the same as for the dark-matter-only runs
described above, except that we add gas particles with zero
gravitational mass. This ensures that the dark matter distribution
remains undisturbed by the inclusion of baryons, so that the halo
merger trees used to generate the semi-analytic galaxy catalogues will
be the same. Gas particles are added at a lower resolution than the
dark matter, simply to ease the computational cost of our
simulations. The resolution we have adopted is sufficient to obtain
numerically converged estimates of bulk cluster properties for systems
with $T\gtrsim 2$ keV (SHT09).

Every time an output redshift is reached in our hydrodynamical simulations,
temporary `galaxy' particles are introduced at positions specified by the
\SAM\ galaxy catalogue. For each galaxy, we know the increase in stellar mass
since the last output, and we remove this mass from the hot phase by converting
the $\Delta N_{\rm star}=\Delta M_*/m_{\rm gas}$ nearest gas particles into
collisionless star particles, using a stochastic method to ensure that $\Delta
N_{\rm star}$ is an integer. Once star formation is complete, we then distribute
metals and the heat energy available from SNe and AGN amongst neighbouring gas
particles in some way, as described in the following sections. Following the
injection of metals and entropy, the galaxy particles are removed and the
simulation continues until the next output time, when the process is
repeated. The main purpose of this paper is to investigate different ways of
heating and enriching intracluster gas, using X-ray observations of
galaxy clusters to constrain our models.

Cluster catalogues are generated at $z=0$ from our simulations using a procedure
similar to that of \citet{MTK02}. Full details of our cluster extraction method
are given in STY10.

Following SHT09, we choose to neglect gas cooling processes in our
hydrodynamical simulations throughout most of this work. Although cooling is
relatively unimportant for the majority of the ICM, we cannot expect to
reproduce the low central entropy and steep entropy profiles of observed CC
clusters, as demonstrated by STY10. However, in Section \ref{sec:cooling} of
this paper, we make a first attempt to overcome this limitation of the model of
SHT09 by including metal-dependent radiative cooling in our simulations. With
the addition of cooling, we show that it is indeed possible to produce both CC
and NCC systems using our hybrid approach.

\section{Feedback from Type II supernovae}
\label{sec:sne}

In this section we investigate how galactic winds driven by Type II SNe shape
the chemical and thermal properties of intracluster gas. We pay particular
attention to how our results are affected by varying the feedback scheme in our
simulations.  Sections~\ref{sec:snmodels}--\ref{sec:snname} describe the models;
then in Sections~\ref{sec:snent} and \ref{sec:snmet} respectively, we show that
SNe simply do not provide enough energy to significantly alter the entropy and
metallicity profiles of the ICM.  Most of the metals in the ICM originate
outside the central cluster galaxy and we argue that the metallicity profile in
these models is imposed by the accretion history of the ICM.

\subsection{Supernova feedback models}
\label{sec:snmodels}

There are two broad classes of SN feedback models deployed in numerical
simulations: {\it thermal}, where the available energy is used to raise the
temperature of neighbouring gas particles
\citep{KAT92,MYT97,THC00,KPF02,BKG04,SSK06}, and {\it kinetic}, where
neighbouring particles are given a velocity `kick'
\citep{NAW93,MIH94,KAW01,KPF02,SPH03,OPD06,DUT08}.

It is well known that simple thermal feedback schemes fail in simulations with
cooling since the injected energy is radiated away before it has any
hydrodynamical effect. This problem is typically evaded by suppressing radiative
cooling by hand. However, this is not an issue for us since cooling processes
are not included in any of our simulations until Section \ref{sec:cooling}. We
have experimented with a variety of both thermal and kinetic models, which we
now describe.

\subsubsection{Thermal models}
\label{sec:thermsnmodels}

The thermal feedback models employed in our simulations can be grouped into
three categories, depending on the method used to inject the SN energy, $\Delta
E_{\rm SN}$, into the ICM.

In our first scheme, we simply heat a fixed number of the gas particles closest
to each galaxy, where the number of neighbours heated is $N_{\rm heat}=1$, $10$
or $100$. This is the approach typically adopted in fully self-consistent
hydrodynamical simulations with radiative cooling, star formation and thermal SN
feedback.

The second method we have investigated is to heat all gas particles within some
sphere centred on each galaxy, where the radius of the sphere is assumed to be
some fraction, $f_{\rm rad}$, of the halo virial radius, $r_{200}$. We have
explored $f_{\rm rad}=0.1$, $0.32$ and $1$. In this model, $N_{\rm heat}$ is
defined as the number of gas particles enclosed by the sphere. If no neighbours
are found, the radius of the sphere is increased until a single gas particle is
found.

Our third approach is to heat neighbouring gas particles by a multiple, $f_{\rm
  temp}$, of the halo virial temperature, $T_{200}$, defined by
\begin{equation}
\label{eq:T200}
T_{200}=\frac{G}{2}\frac{\mu m_{\rm H}}{k_{\rm B}}\frac{M_{200}}{r_{200}}.
\end{equation}
The values of $f_{\rm temp}$ that we adopt are $1$, $3.2$ and
$10$. The number of gas particles that can be heated with the available energy
is then
\begin{equation}
N_{\rm heat}=\frac{\mu m_{\rm H}(\gamma-1)\Delta E_{\rm SN}}{f_{\rm
    temp}k_{\rm B}T_{200}m_{\rm gas}}.
\end{equation}

In GADGET-2 the thermodynamic state of each fluid element is defined in terms of
the entropic function
\begin{equation}
A_i=\frac{(\gamma -1)u_i}{\rho_i^{\gamma-1}},
\end{equation}
where $u_i$ is the thermal energy per unit mass of a particle and $\rho_i$ is
its density. Supplying heat energy to a gas particle causes $A_i$ to
increase. Note that $A$ is related to the X-ray gas entropy $K$ via $K=\mu
m_{\rm H}(\mu_{\rm e}m_{\rm H})^{\gamma-1}A$, where $\mu_{\rm
  e}m_\mathrm{H}\approx1.90\times10^{-27}$\,kg is the mean molecular mass per
free electron.

In each of our three feedback schemes, we heat particles by raising their
thermal energy by a fixed amount
\begin{equation}
\label{eq:du}
\Delta u_i = \frac{\Delta E_{\rm SN}}{N_{\rm heat}m_{\rm gas}},
\end{equation} 
implemented in GADGET-2 as an entropy boost of
\begin{equation}
\label{eq:dAE}
\Delta A_i = \frac{(\gamma-1)\Delta u_i}{[\max{(f_{\rm b}\rho_{\rm 200},\rho_i)}]^{\gamma-1}}.
\end{equation}
The product of the cosmic baryon fraction, $f_{\rm b}$, and the virial density,
$\rho_{200}$, gives the mean overdensity of baryons within the virial radius. If
the required $N_{\rm heat}$ neighbours are not found within a distance $r_{200}$
of a galaxy and the search radius has to be increased, the density of some of
these particles may be less than $f_{\rm b}\rho_{200}$. By using $[\max{(f_{\rm
      b}\rho_{200},\rho_i)}]^{\gamma-1}$, rather than $\rho_i^{\gamma-1}$, in
the denominator of equation (\ref{eq:dAE}), we are assuming that the amount of
energy used to heat such particles is $\Delta u_i(\rho_i/f_{\rm
  b}\rho_{200})^{\gamma-1}<\Delta u_i$; the rest of the energy is taken to be
used up as the gas does work expanding adiabatically to a density $\rho_i<f_{\rm
  b}\rho_{200}$.

For the first two schemes mentioned above, we have also tested an alternative
heating model where gas particles are given a fixed entropy, rather than energy,
boost:
\begin{equation}
\label{eq:dA}
\Delta A_i=\frac{(\gamma-1)N_{\rm heat}\Delta u_{i}}{\sum_{j=1}^{N_{\rm
      heat}}[\max{(f_{\rm b}\rho_{\rm 200},\rho_j)}]^{\gamma-1}}.
\end{equation}
Denser particles close to a galaxy are then heated to a higher temperature than
more distant, lower density particles.

\subsubsection{Kinetic models}
\label{sec:kinsnmodels}

Kinetic SN feedback is implemented in our simulations by assuming that
gas particles closest to a model galaxy are given a velocity kick. The number of
particles that receive a kick depends on the available energy:
\begin{equation}
\label{eq:nkick}
N_{\rm kick}=\frac{2\Delta E_{\rm SN}}{m_{\rm gas}v_{\rm wind}^2},
\end{equation}
where the wind speed, $v_{\rm wind}$, is a free parameter. We have considered
several different values for the wind speed: $v_{\rm wind}=1$ km s$^{-1}$, $300$
km s$^{-1}$, $600$ km s$^{-1}$ and $1000$ km s$^{-1}$. For any given galaxy, we
impose the constraint that the wind speed cannot be less than the virial
speed\footnote{We define the virial speed of a halo to be the circular velocity
  at the virial radius.} of the host halo, $v_{200}$, so that the case $v_{\rm
  wind}=1$ km s$^{-1}$ is equivalent to assuming that material is ejected at the
virial speed. To ensure that $N_{\rm kick}$ is an integer, we draw a random
number $r$ uniformly from the unit interval and compare it with the fractional
part of $N_{\rm kick}$: if $r$ is less (greater) than the fractional part of
$N_{\rm kick}$, we round $N_{\rm kick}$ up (down) to the nearest integer.

The velocity of each kicked particle is modified according to
\begin{equation}
\mathbf{v}\rightarrow\mathbf{v}+v_{\rm wind}\mathbf{\hat{n}},
\end{equation}
where $\mathbf{\hat{n}}$ is a unit vector that is either oriented in a random
direction on the unit sphere, or in the direction from the galaxy to the wind
particle.

We have only studied kinetic feedback models where the wind speed is a constant
for all galaxies. Similar models are often employed in self-consistent
hydrodynamical simulations (e.g. \citealt{NAW93,SPH03,DVS08}). However, there
are other possibilities, such as momentum-driven winds, where the wind speed
scales with the galaxy velocity dispersion (e.g. \citealt{MAR05,OPD06}), and
models where the outflow velocity increases with galactocentric radius
\citep{SES10}.

\subsection{Metal enrichment models}
\label{sec:Zmodels}

We distribute metals amongst gas particles in our simulations as follows. For
each model galaxy, all gas particles contained within a sphere centred on the
galaxy are identified. As before, the radius of the sphere is chosen to be a
fraction, $f_{\rm Z,rad}$, of the halo virial radius, where $f_{\rm Z,rad}=0.1$,
$0.32$ or $1$. The metals in diffuse gas produced by the galaxy are then shared
evenly amongst these particles, so that the metal mass associated with each
particle, $m_{{\rm Z},i}$, increases by an amount
\begin{equation}
\Delta m_{\rm Z} = \frac{\Delta M_{\rm Z,ICM}}{N_{\rm enrich}},
\end{equation}
where $N_{\rm enrich}$ is the number of gas particles inside the sphere. 

Given the total mass in metals for a gas particle, we could then define its
metallicity simply as
\begin{equation}
Z_{{\rm part},i}=\frac{m_{{\rm Z},i}}{m_{\rm gas}},
\end{equation}
which we refer to as the \emph{particle metallicity}. However, in this work, we
prefer to use the \emph{smoothed metallicity} \citep{OEF05,TBD07}, defined by
\begin{equation}
\label{eq:Zsm}
Z_{{\rm sm},i}=\frac{\rho_{{\rm Z},i}}{\rho_{i}},
\end{equation}
where the smoothed metal mass density, $\rho_{{\rm Z},i}$, is computed in an
analogous way to the standard SPH density estimate:
\begin{equation}
\rho_{{\rm Z},i}=\sum_{j=1}^{N_{\rm sph}}m_{{\rm Z},j}W(|\mathbf{r}_i-\mathbf{r}_j|,h_i).
\end{equation}
Here $N_{\rm sph}=64$ is the number of SPH smoothing neighbours and $W$ is a
spherically-symmetric smoothing kernel, which depends upon the separation of
particles $i$ and $j$, $|\mathbf{r}_i-\mathbf{r}_j|$, and the smoothing length
of particle $i$, $h_i$. The smoothed metallicity of a gas particle is updated
whenever its SPH density is calculated. Once a gas particle is converted into a
star particle, its smoothed metallicity remains fixed for the rest of the
simulation. Note that the metallicity of particles does not affect the gas
dynamics in our \changed{non-radiative} simulations.

\subsection{Naming conventions}
\label{sec:snname}

Table \ref{tab:sn2models} lists all $23$ of our SN feedback models.  Note that
each model, including the reference gravitational heating only model (GO),
follows the conversion of gas into stars as dictated by the underlying SA model.

\begin{table*}
\begin{minipage}{126mm}
\caption{Supernova feedback models.  Unless otherwise stated, the radius for
  metal injection is $r_{200}$ ($f_{\rm Z,rad}=1$).  Note that all models,
  including the GO model, follow the conversion of gas into stars.}
\label{tab:sn2models}
\begin{tabular}{@{}llll}
\hline
Model name & Type & Energy injection method & Comments \\
\hline
GO & - & - & Gravitational heating only \\
\hline
SN\_Th\_N$N_\mathrm{heat}$ & Thermal & Fixed energy & $N_{\rm heat}=1$, $10$, $100$\\
SN\_Th\_R$f_{\rm rad}$ & Thermal & Fixed energy & $f_{\rm rad}=0.1$, $0.32$, $1$\\
SN\_Th\_T$f_{\rm temp}$ & Thermal & Fixed energy & $f_{\rm temp}=1$, $3.2$, $10$\\
\hline
SN\_En\_N$N_{\rm heat}$ & Thermal & Fixed entropy & $N_{\rm heat}=1$, $10$, $100$\\
SN\_En\_R$f_{\rm rad}$ & Thermal & Fixed entropy & $f_{\rm rad}=0.1$, $0.32$, $1$\\
\hline
SN\_KiR\_V$v_{\rm wind}$ & Kinetic & Velocity kick, random & $v_{\rm
  wind}$/km\,s$^{-1}=1$, 300, 600, 1000\\
SN\_KiD\_V$v_{\rm wind}$ & Kinetic & Velocity kick, directed & $v_{\rm wind}$/km\,s$^{-1}=600$\\
\hline
SN\_KiR\_Z$f_{\rm Z,rad}$ & Kinetic & Velocity kick, random & $f_{\rm
  Z,rad}=0.1$, $0.32$, $1$; $v_{\rm wind}$/km\,s$^{-1}=600$ \\
\hline
\end{tabular}
\end{minipage}
\end{table*}

\subsection{Entropy profiles}
\label{sec:snent}

\begin{figure}
\includegraphics[width=85mm]{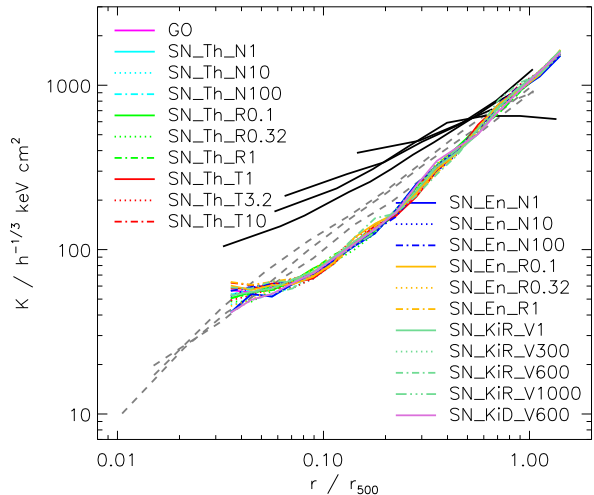}
\caption{Entropy profiles for cluster C1 resimulated with different
  implementations of supernova feedback (coloured lines; see legend for model
  details). Note that the profile obtained from a gravitational heating only
  model (GO) is also shown. For comparison, we display observed profiles of
  similar-mass CC (dashed grey lines) and NCC (solid black lines) clusters in
  the REXCESS sample (PAP10).}
\label{fig:snKprof}
\end{figure}

To test the effect of different implementations of SN feedback
on the entropy structure of the ICM, we have resimulated our fiducial
cluster, C1, with each of our models. Figure \ref{fig:snKprof} shows
the resulting entropy profiles. For comparison, we also show the
observed entropy profiles of CC and NCC clusters in the REXCESS sample
\citep[hereafter PAP10]{PAP10} . To facilitate a fair comparison, we
only plot profiles of observed clusters that have a mass,
$M_{500}$, within $20\%$ of that of cluster C1.

The main point to note is that all of our models yield almost identical entropy
profiles that are in good agreement with the profile obtained from the
reference GO run. In all cases, the profiles scale approximately as $K\propto
r^{1.2}$ for $r\gtrsim 0.1r_{500}$, consistent with spherical accretion models
(e.g. \citealt{TON01}) and cosmological simulations that include gravitational
heating only (e.g. \citealt{VKB05, NKV07}).  For $r\lesssim 0.1r_{500}$, the
entropy profiles flatten off significantly, exhibiting a small spread in central
entropy. 

Compared to the observed entropy profiles of CC clusters, the profiles predicted
by our models have a steeper slope at $r\gtrsim 0.1r_{500}$, and the
normalisation is systematically too low. In the case of NCC clusters, it is
evident that none of our models can explain the shallow profiles characteristic
of these systems.

We have checked that these results hold for other clusters in our sample, so we
conclude that SNe have a negligible impact on the thermodynamical properties of
intracluster gas and, furthermore, the manner in which the feedback energy is
injected is unimportant.

\subsection{Metallicity profiles}
\label{sec:snmet}

\begin{figure}
\includegraphics[width=85mm]{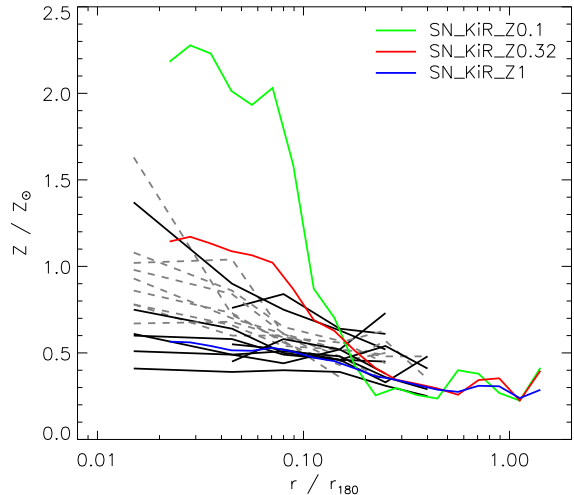}
\caption{Emission-weighted metallicity profiles for cluster C1
  resimulated with the same kinetic supernova feedback model, but
  varying the radius of the region within which metals are injected
  (solid coloured lines). See the legend for details of the metal
  enrichment models adopted. Observed profiles of CC (dashed grey
  lines) and NCC (solid black lines) clusters from the sample of MAT11
  are also shown.}
\label{fig:snZprof}
\end{figure}

The metal enrichment model has only one free parameter, $f_{\rm
  Z,rad}$, which controls the radius of the spherical region about a
galaxy in which metals are injected. We have resimulated cluster C1
with three different values of $f_{\rm Z,rad}$, fixing the SN feedback
scheme to be the kinetic model where gas particles are kicked in a
random direction with $v_{\rm wind}=600$ km s$^{-1}$. Figure
\ref{fig:snZprof} shows the emission-weighted metallicity profiles
that result, along with observed Fe abundance profiles of CC and NCC
clusters from \citet[hereafter MAT11]{MAT11}. We plot all observed
clusters with a mass above $80\%$ of that of cluster C1, in order to
obtain a reasonable number of both CC and NCC objects (most NCC
clusters in the sample of MAT11 are considerably more massive than
cluster C1).

We note that it is difficult to directly compare the metallicity
profiles of our clusters with those of observed clusters from MAT11, for several
reasons. Firstly, the observed profiles are Fe abundance profiles, but our
simple metal enrichment model does not include the contribution from Type 1a
SNe, a major source of Fe, nor does it track the production of individual
chemical elements. Secondly, we could, in principle, adjust the yield in the SA
model underpinning our simulations, which would allow us to alter the
normalisation of our metallicity profiles. For these reasons, we focus on the
shape of metallicity profiles, instead of their normalisation, when assessing
the impact of different feedback and enrichment models on the ICM enrichment
pattern.


It is apparent from Figure \ref{fig:snZprof} that varying the metal
injection radius has a large impact on ICM metallicity profiles in
core regions, $r\lesssim 0.2r_{180}$. If metals are injected in a
concentrated fashion ($f_{\rm Z,rad}=0.1$), then we see a sharp peak
in the metal distribution within that region that is not reflected in
the observational data. As the injection radius is increased, the
gradient of the profile becomes progressively shallower until, when
$f_{\rm Z,rad}=1$, the slope provides a good match to that of the
profiles of observed NCC clusters, except the few systems that have a
high central abundance more typically found in CC clusters. It is
possible that these systems are CC remnants. Recall that radiative
cooling is not included in our simulations so we do not expect to be
able to reproduce the abundance peaks seen in the core regions of CC
clusters.

\begin{figure}
\includegraphics[width=85mm]{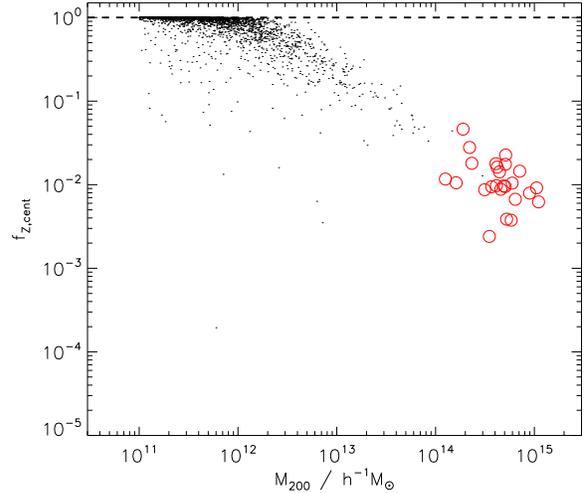}
\caption{Fraction of metals in hot halo gas produced by the central galaxy of
  the halo, $f_{\rm Z,cent}$, as a function of virial mass. The black points are
  for a subsample of halos with $M_{200}\geq 10^{11}h^{-1}\Msun$ extracted from
  the Millennium Simulation galaxy catalogues. The red circles correspond to our
  $25$ resimulated clusters. For these massive systems, under $5\%$ of metals in
  the hot gas are produced by the central galaxy, implying that nearly all of
  the metals are accreted.}
\label{fig:metself}
\end{figure}

A metal enrichment model where metals are distributed throughout the
halo can be justified if the bulk of the metals found in intracluster
gas were brought in by infalling material, rather than being produced
by star formation in the central galaxy of the halo.  To check whether
this is the case, we have modified the \lgalaxies\ \SAM\ to follow
what fraction of metals in diffuse halo gas are produced by the
central galaxy of the halo. Figure \ref{fig:metself} shows this
fraction as a function of halo virial mass for all $25$ clusters in
our sample, and for a selection of halos with $M_{200}\geq
10^{11}h^{-1}\Msun$ taken from the Millennium Simulation galaxy
catalogues. For all of our clusters, the fraction of metals in hot
halo gas produced by the central galaxy is less than $5\%$, implying
that nearly all of the metals in diffuse gas are indeed accreted. Note
that this may change somewhat when we extend \lgalaxies\ to track the
time-dependence of metals returned by Type Ia SNe and AGB stars, as
some of the metal production will be delayed until after the formation
of the central cluster galaxy.

Ideally one would like to inject metals locally about satellite galaxies falling
into a halo, rather than distributing them uniformly throughout the
halo. However, this is not possible with the \citet{DLB07} version of
\lgalaxies\ since all the metals in the diffuse gas associated with a galaxy are
assumed to be instantaneously stripped once it becomes a satellite galaxy. In
future work we plan to switch to a different treatment of satellite galaxies
whereby hot gas is gradually removed from infalling galaxies by tidal and
ram-pressure stripping \citep[e.g.~][]{HET10}.

\begin{figure}
\includegraphics[width=85mm]{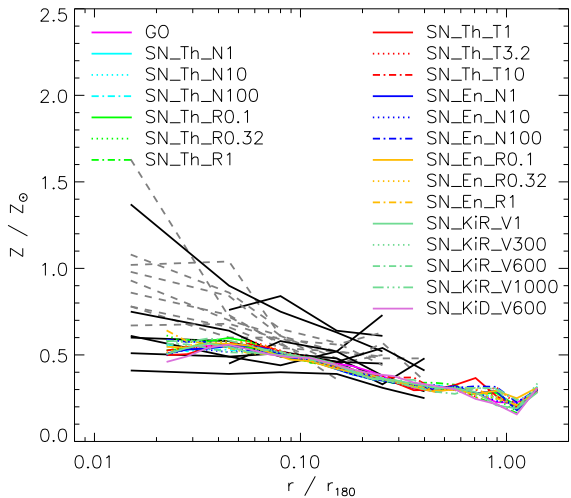}
\caption{Emission-weighted metallicity profiles for cluster C1
  resimulated with different supernova feedback schemes, assuming that
  metals are distributed uniformly throughout the halo (coloured
  lines; see the legend for feedback model details). We also show
  profiles of CC (dashed grey lines) and NCC (solid black lines)
  clusters from the observational sample of MAT11.}
\label{fig:allsnZprof}
\end{figure}

We now examine whether changing the SN feedback scheme affects ICM
metallicity profiles. To do this, we have resimulated our fiducial
cluster with all of the models described in Section
\ref{sec:snmodels}, fixing $f_{\rm Z,rad}=1$ in each case. Figure
\ref{fig:allsnZprof} compares the emission-weighted metallicity
profiles obtained from our cluster simulations with those of the same
observed clusters from the MAT11 sample.

The main  point to note is that  the metallicity profiles obtained  from all our
various  simulations are  essentially the  same and  we have  checked  that this
conclusion  remains valid  when metals  are  injected in  a concentrated  manner
($f_{\rm Z,rad}=0.1$).  It  follows that SN feedback has no  impact on the metal
distribution in clusters, and the precise way in which the energy available from
SNe is used to heat intracluster gas is irrelevant.

\changed{The metallicity profiles of clusters in the presence of
  supernova feedback were investigated by \citet{TBD07}.  They also
  found that changing the SNe feedback rate makes little difference to
  the slope of the profiles (although it does change the
  normalisation).  We show in Figure~\ref{fig:agnZprof} and
  Section~\ref{sec:agnmet} that the stronger AGN jet feedback can have
  a larger effect.}


\subsection{Summary}

Our study so far has revealed that feedback from SNe has a negligible effect on
both ICM entropy and metallicity profiles, regardless of the manner in which the
energy is assumed to be transferred to the gas.

In light of this freedom, we choose our fiducial SN feedback scheme to be the
kinetic model SN\_KiR\_V600, where gas particles are given a kick in a random
direction with $v_{\rm wind}=600$ km s$^{-1}$. This is similar to the model of
\citet{DVS08}. A wind speed of $600$ km s$^{-1}$ is consistent with observations
of local (e.g. \citealt{VCB05}) and $z\sim2$--3 (e.g. \citealt{SES10}) starburst
galaxies.

The metallicity profiles reflect the manner in which metals are injected into
the diffuse gas. For our fiducial metal enrichment model we assume that the
metals ejected from galaxies are distributed uniformly throughout the entire
halo, since this gives a good match to the slope of the metallicity profiles of
observed NCC clusters. This model is justified by the fact that nearly all of
the metals in intracluster gas are accreted, rather than being produced by the
central galaxy of the halo.

The model that forms the basis for the rest of the work presented in this paper
is thus SN\_KiR\_Z1.

\section{Feedback from active galactic nuclei}
\label{sec:agn}

As shown in the previous section, the heating of intracluster gas by stellar
feedback alone clearly cannot account for the excess entropy observed in cluster
cores, indicating that an additional feedback mechanism must be at play. The
favoured candidate is the energy liberated by the accretion of gas onto central
supermassive black holes at the centres of galaxies. Our goal in this section is
to assess how the properties of the ICM are altered by the inclusion of this
extra heating from AGN, and how our results are affected by different numerical
implementations of AGN feedback.

In Section~\ref{sec:agnmodels} we describe simple AGN heating models, similar to
those found in the literature, then in Section~\ref{sec:agnent} we use
comparisons with observed entropy profiles to conclude that none of these models
are entirely satisfactory.  One model with extreme wind speeds does provide an
adequate fit to the data and that motivates the new stochastic heating model
developed in Section~\ref{sec:newagn}.

\subsection{AGN feedback models}
\label{sec:agnmodels}

\changed{The amount of energy available from AGN heating, $\Delta E_{\rm AGN}$,
  is not arbitrary but is set by the model described in Secton~3.1.2 of
  SHT09.  Although that paper considered only a single heating model, we
  find that the global X-ray lumnosity-temperature relation is dependent mainly
  upon the normalisation of the heating, and is relatively unaffected by the
  particular manner in which the heat is injected.  That can have a large effect
  on the entropy profiles, however, as we show below.}

\subsubsection{Thermal models}

The first set of thermal AGN feedback models that we have tested are identical
to the thermal SN feedback models described previously in Section
\ref{sec:thermsnmodels}, except with $\Delta E_{\rm SN}$ replaced by $\Delta
E_{\rm AGN}$. Similar prescriptions for AGN feedback have been employed in
numerous other works (e.g. \citealt{SDH05a,DCS08,BOS09,FBT10}).  

\subsubsection{Kinetic models}

We implement kinetic AGN feedback in our simulations in the same way as kinetic
SN feedback; see Section \ref{sec:kinsnmodels}. The only differences are
that the number of particles kicked (equation \ref{eq:nkick}) now depends on the
energy available from black hole accretion, $\Delta E_{\rm AGN}$, rather than
that available from SN explosions, $\Delta E_{\rm SN}$, and we have
adopted larger wind speed values, $v_{\rm wind}=1000$ km s$^{-1}$, $4500$ km
s$^{-1}$ and $20000$ km s$^{-1}$, in line with measured AGN outflow velocities
\citep{PRK03,CBG03,CKG03,GAB08,DBA10}.



\subsubsection{Naming conventions}

Table \ref{tab:agnmodels} lists all of our various AGN feedback models. In each
case the SN feedback and metal injection schemes are the same as for model SN\_KiR\_Z1.

\begin{table*}
\begin{minipage}{126mm}
\caption{AGN feedback models. In each case, supernova feedback is implemented
  using a kinetic model where particles neighbouring a galaxy are given a kick
  in a random direction with velocity $600$ km s$^{-1}$, and ejected metals are
  assumed to be distributed uniformly throughout the entire host halo.}
\label{tab:agnmodels}
\begin{tabular}{@{}llll}
\hline
Model name & Type & Energy injection method & Comments \\
\hline
AGN\_Th\_N$N_\mathrm{heat}$ & Thermal & Fixed energy & $N_{\rm heat}=1$, $10$, $100$\\
AGN\_Th\_R$f_{\rm rad}$ & Thermal & Fixed energy & $f_{\rm rad}=0.1$, $0.32$, $1$\\
AGN\_Th\_T$f_{\rm temp}$ & Thermal & Fixed energy & $f_{\rm temp}=1$, $3.2$, $10$\\
\hline
AGN\_En\_N$N_{\rm heat}$ & Thermal & Fixed entropy & $N_{\rm heat}=1$, $10$, $100$\\
AGN\_En\_R$f_{\rm rad}$ & Thermal & Fixed entropy & $f_{\rm rad}=0.1$, $0.32$, $1$\\
\hline
AGN\_KiR\_V$v_{\rm wind}$ & Kinetic & Velocity kick, random & $v_{\rm
  wind}$/km\,s$^{-1}=$ 1000, 4\,500, 20\,000\\
AGN\_KiD\_V$v_{\rm wind}$ & Kinetic & Velocity kick, directed & $v_{\rm
  wind}$/km\,s$^{-1}=$ 20\,000\\ 
\hline
\end{tabular}
\end{minipage}
\end{table*}

\subsection{Entropy profiles}
\label{sec:agnent}

In order to assess how sensitive the thermodynamical properties of the ICM are
to different implementations of AGN feedback, we have resimulated our fiducial
cluster with each of our $15$ thermal and $4$ kinetic AGN feedback models.  

\subsubsection{Thermal models}

\begin{figure}
\includegraphics[width=85mm]{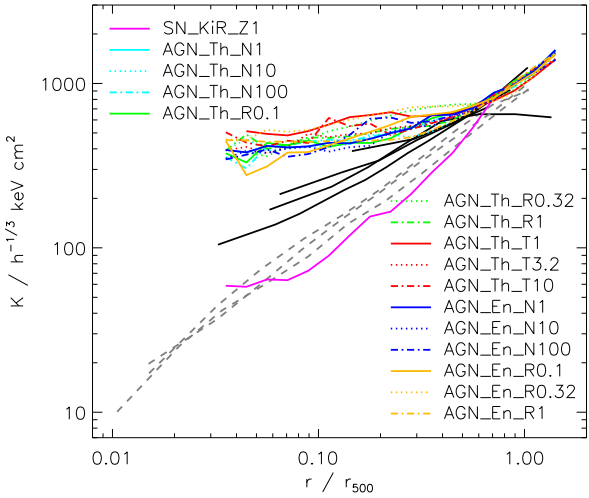}
\caption{Entropy profiles for cluster C1 resimulated with different
  implementations of thermal AGN feedback (coloured lines; see the
  legend for model details). For comparsion, we also show the profile
  obtained from a run with kinetic supernova feedback only, model
  SN\_KiR\_Z1, and observed profiles of CC (dashed grey lines) and NCC
  (solid black lines) clusters in the REXCESS sample (PAP10).}
\label{fig:thermagnKprof}
\end{figure}

The entropy profiles obtained from our thermal models are displayed in Figure
\ref{fig:thermagnKprof}. For comparison, we also show the profile predicted by
our fiducial SN feedback model (SN\_KiR\_Z1), and entropy profiles of observed
clusters of a similar mass.

As for the SN feedback, all of the thermal models
give very similar results, even though there are large differences in the way in
which the available energy is shared amongst gas particles in the various
schemes. Essentially, the profiles follow the predicted $r^{1.1-1.2}$ scaling at
large radii, but as we move in towards the core they begin to flatten off at
$r\sim 0.5-0.6r_{500}$. At radii interior to this, the slope is shallower than
seen in either CC or NCC clusters, leading to an over-estimate of the central
entropy. Simple preheating models predict similarly large isentropic cores at
$z=0$ (e.g. STY10).

\subsubsection{Kinetic models}

\begin{figure}
\includegraphics[width=85mm]{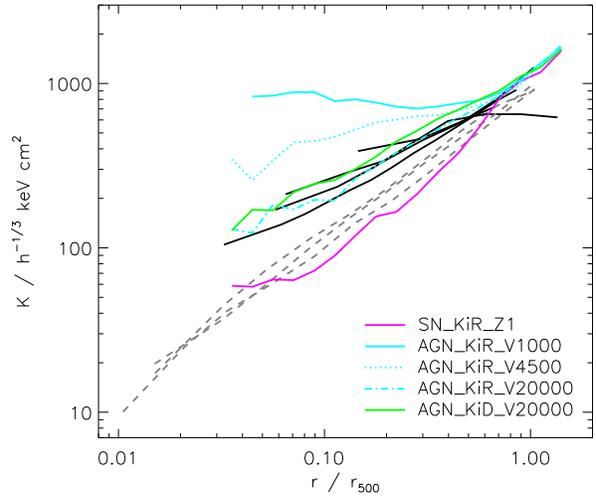}
\caption{Entropy profiles for cluster C1 resimulated with different kinetic AGN
  feedback models (coloured lines; see legend for model details). For
  comparsion, we also show the profile obtained from a run with kinetic
  supernova feedback only, model SN\_KiR\_Z1, and observed profiles of CC
  (dashed grey lines) and NCC (solid black lines) clusters in the REXCESS sample
  (PAP10).}
\label{fig:kinagnKprof}
\end{figure}

Figure \ref{fig:kinagnKprof} compares the entropy profiles predicted by our
kinetic AGN feedback models with the same set of observed cluster profiles as in
Figure \ref{fig:thermagnKprof}.

In cluster outskirts, $r\sim r_{500}$, the models give similar
results, but there are clear difference at radii less than this. For
the lowest wind speed, $v_{\rm wind}=1000$ km s$^{-1}$, we see a very
flat entropy profile, with a hint of an entropy inversion in the
core. As the wind speed is increased, the entropy profile steadily
steepens, providing a good match to observed NCC cluster profiles when
$v_{\rm wind}=20\,000$ km s$^{-1}$.  Almost identical results are
obtained when kicks are imposed in the direction from the galaxy to
the wind particle, rather than in a random direction.

To understand this behaviour, we have checked how the trajectories of kicked
particles are affected by variations in the wind speed. This is done by
identifying the main progenitor of our cluster at high-redshift using the halo
merger trees, selecting all gas particles within $r_{500}$ of this object that
have just received a kick, then tracking the cluster-centric positions of these
particles to $z=0$. For a high wind speed, $v_{\rm wind}=20\,000$ km s$^{-1}$, the
available AGN energy is only sufficient to kick a small number of particles and
we find that their large momentum carries them beyond $r_{500}$. As we reduce
the wind speed to $v_{\rm wind}=1000$ km s$^{-1}$, the number of particles
kicked increases but their momentum gain is smaller, so they do not escape from
the cluster core before their kinetic energy is converted to thermal
energy. This leads to an increase of the gas entropy in the central regions,
establishing a flat entropy profile as seen in Figure \ref{fig:kinagnKprof}.

\subsection{Summary}

Several interesting results have emerged from our study of the effect of
different AGN feedback models on the thermal properties of the ICM.

We have found that simple thermal feedback schemes, based on heating a fixed
number of particles, heating particles within a fixed fraction of the virial
radius, or heating particles by a fixed fraction of the virial temperature, all
heat the gas in cluster central regions excessively, leading to a higher core
entropy than observed. All $15$ of the thermal models we have tested give very
similar results.

When AGN feedback is implemented in a kinetic manner, ICM entropy profiles are
found to be sensitive to the wind speed adopted. For low wind speeds, the
resulting entropy profiles are too flat, as in the thermal case. This is because
the available energy is shared amongst a large number of particles, and kicked
particles do not have sufficient momentum to escape central cluster regions
before their kinetic energy is thermalised.

As the wind speed is increased, the number of particles kicked decreases as
$1/v_{\rm wind}^2$ and kicked particles are able to reach larger cluster-centric
radii before thermalisation of their kinetic energy. Consequently, more
low-entropy material remains in core regions, so entropy profiles become
progressively steeper, approaching observed ones. For $v_{\rm wind}=20\,000$ km
s$^{-1}$, the predicted profiles agree well with observed profiles of NCC
clusters. However, such a high wind speed is perhaps physically unrealistic.

From our discussion, it seems that the key ingredient of a successful AGN
feedback model must be to ensure that only a small fraction of particles in
central cluster regions are heated/kicked, so that these particles have
sufficient entropy/momentum to reach cluster outskirts, leaving low-entropy
metal-rich gas behind in the core. In the next section we formulate a new AGN
feedback prescription that has this desired feature, and is motivated by the
observed interaction of AGN with their environment.

\section{A new model for feedback from active galactic nuclei}
\label{sec:newagn}

There is a growing body of observational evidence that AGN feedback may be
mostly related to radio-loud AGN. In the local universe, observations of galaxy
groups and clusters often show X-ray cavities coincident with lobes of radio
emission linked to the central galaxy by radio jets
\citep{BSM01,BRM04,MNW05,FST06,MIY06,JHP08,GBT09,DRM10,GOV11}. It is thought
that these bubbles are inflated by the central AGN, and may provide an efficient
means of removing cool, enriched gas from cluster cores as they rise buoyantly
through the cluster atmosphere, thus quenching star formation.

At high redshift, $z\sim 2-3$, emission-line kinematics of radio galaxies based
on rest-frame optical integral-field spectroscopy have revealed powerful bipolar
outflows with kinetic energies equivalent to $0.2\%$ of the rest mass of the
central supermassive black hole (e.g. \citealt{NLE06,NLD08}). These AGN-driven
winds are energetic enough to remove copious amounts of gas from the host
galaxy, preventing further accretion onto the black hole and suppressing star
formation. Large-scale energetic outflows have also been observed in $z\approx
2$ ultra-luminous infrared galaxies \citep{ASS10}, a galaxy population
potentially an order of magnitude more common than distant radio galaxies.

Although it is not yet fully understood how the energy released by black hole
accretion is transferred to the surrounding gas, the observational data suggests
that the energy is input in a directional manner, via jets or collimated
outflows, rather than isotropically. To reflect this, we have developed an
anisotropic, stochastic heating model where only some of the gas particles
neighbouring a galaxy are heated per duty cycle of the AGN.  We note
that higher-resolution models of feedback from AGN in cluster cores
also favour anisotropic heating \citep[e.g.~][]{GRS12}.

In Section~\ref{sec:newagnmodel} we describe this heating model in detail, then
in Section~\ref{sec:newagnent} we use observed entropy profiles and
  X-ray scaling relations to determine optimal model parameters.

\subsection{Stochastic AGN feedback model}
\label{sec:newagnmodel}

The basis of our new model for AGN heating is as follows. For each galaxy, we
first identify all gas particles contained within a sphere centred on the
galaxy, where the radius of the sphere is some fraction, $f_{\rm rad}$, of the
halo virial radius. We then assume that the probability that any of these
particles has been heated by AGN feedback during the time elapsed, $\Delta t$,
since the previous \SAM\ output is
\begin{equation}
P_{\rm heat}=1-(1-f_{\rm duty})^{\Delta t/t_{\rm duty}},
\end{equation}
where $f_{\rm duty}$ is a parameter controlling the fraction of particles heated
over the AGN duty cycle, $t_{\rm duty}$. Based on observational data, we take
$t_{\rm duty}=10^8$ yrs (e.g. \citealt{BRM04,FST06,JHP08}). With our choice of
\SAM\ output times we then have $2\lesssim \Delta t/t_{\rm duty}\lesssim 4$ for
$z<3$.

For each gas particle neighbour, we draw a random number $r$ uniformly from the
unit interval and compare it with $P_{\rm heat}$: if $r<P_{\rm heat}$ the
particle is given an entropy boost
\begin{equation}
\label{eq:stochasticdA}
\Delta A_i=\frac{(\gamma-1)\Delta E_{\rm AGN}}{m_{\rm gas}P_{\rm
    heat}\sum_{j=1}^{N_{\rm heat}}\left[\max{(f_{\rm
        b}\rho_{200},\rho_j)}\right]^{\gamma-1}},
\end{equation}
and if $r\geq P_{\rm heat}$ the particle is not heated. By including the heating
probability $P_{\rm heat}$ in the denominator of equation
(\ref{eq:stochasticdA}), we ensure that the total amount of energy injected into
the gas is (approximately) the same for different choices of $f_{\rm duty}$.

We have also experimented with supplying the AGN heat energy to particles as a
fixed energy boost. However, this makes virtually no difference to our results
so we do not discuss these models hereafter.

There are two free parameters in our model: $f_{\rm rad}$ and $f_{\rm
  duty}$. The values of these parameters we have tested in this work are $f_{\rm
  rad}=0.1$, $0.32$ and $1$, and $f_{\rm duty}=10^{-4}$, $10^{-3}$, $10^{-2}$
and $10^{-1}$. Note that in the case $f_{\rm duty}=1$ our model reduces to
AGN\_Th\_R$f_{\rm rad}$. 

It is interesting to link the parameter $f_{\rm duty}$ to the opening angle of
AGN jets. If we make the simple approximation that large-scale AGN outflows can
be treated as biconical jets, each with opening angle $2\theta$, then it follows
that $\cos{\theta}=1-f_{\rm duty}$. For the range of values of $f_{\rm duty}$
tested here, this corresponds to $1^{\circ}\lesssim\theta\lesssim 26^{\circ}$.

Table \ref{tab:agnmodels} lists all of our stochastic AGN feedback models.  To
distinguish these from the AGN heating models of the previous section,
we have given them the label JET. 

\begin{table*}
\begin{minipage}{126mm}
\caption{Stochastic AGN feedback models. In each case, supernova feedback is
  implemented using a kinetic model where particles neighbouring a galaxy are
  given a kick in a random direction with velocity $600$ km s$^{-1}$. Unless
  otherwise stated, the radius for energy and metal injection is $r_{200}$
  ($f_{\rm rad}=1$ and $f_{\rm Z,rad}=1$, respectively), and the fraction of
  particles heated per AGN duty cycle is $f_{\rm duty}=10^{-2}$.}
\label{tab:newagnmodels}
\begin{tabular}{@{}llll}
\hline
Model name & Type & Energy injection method & Comments \\
\hline
JET\_R$f_\mathrm{rad}$\_D$f_{\rm duty}$ & Stochastic & Fixed entropy & $f_{\rm rad}=0.1$, $0.32$, $1$\\
& & & $f_{\rm duty}=10^{-4}$, $10^{-3}$, $10^{-2}$, $10^{-1}$\\
\hline
JET\_Z$f_{\rm Z,rad}$ & Stochastic & Fixed entropy & $f_{\rm Z,rad}=0.1$, $0.32$, $1$\\
\hline
\end{tabular}
\end{minipage}
\end{table*}

\subsection{Entropy profiles and scaling relations}
\label{sec:newagnent}

We now investigate whether our new physically-motivated stochastic AGN feedback
scheme yields a better match to observed cluster profiles than the simple
thermal and kinetic models discussed in the previous section.

Recall that our stochastic model has two free parameters: $f_{\rm rad}$, which
governs the radius about a galaxy in which energy is injected, and the fraction,
$f_{\rm duty}$, of neighbouring gas particles that are heated per AGN duty
cycle. The first issue to address is how varying these parameters affects
cluster properties. We then identify an optimal choice for these parameters by
using a selection of observational data to constrain our model.

\subsubsection{The effect of changing $f_{\rm rad}$}

\begin{figure}
\includegraphics[width=85mm]{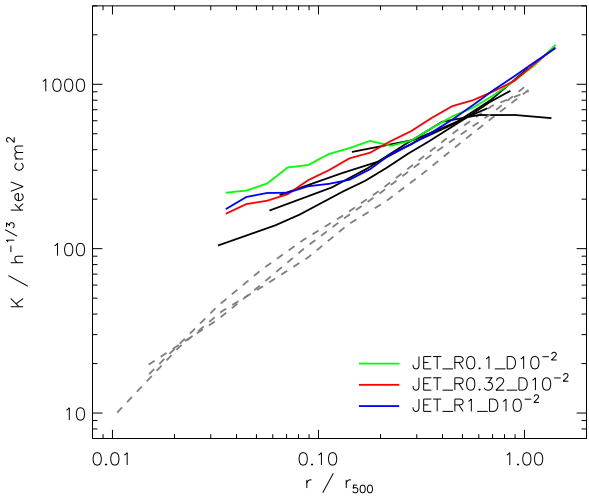}
\caption{Entropy profiles for cluster C1 resimulated with our stochastic AGN
  feedback model for different values of the parameter $f_{\rm rad}$ (solid
  coloured lines; see legend for model details). The other parameter in the
  model, $f_{\rm duty}$, is fixed at $10^{-2}$. The profiles of observed CC
  (dashed grey lines) and NCC (solid black lines) clusters in the REXCESS sample
  (PAP10) are also displayed for comparison.}
\label{fig:fradvary}
\end{figure}

Figure \ref{fig:fradvary} shows the effect of varying $f_{\rm rad}$ on the entropy
profile of cluster C1. In each case, $f_{\rm duty}$ is kept fixed at
$10^{-2}$. It is apparent that the entropy structure of the ICM is relatively
insensitive to the choice of $f_{\rm rad}$, with only small differences between
the three different runs. The best match to observed NCC cluster profiles arises
when $f_{\rm rad}=1$, in which case we find excellent agreement with the
observational data.

\begin{figure}
\includegraphics[width=85mm]{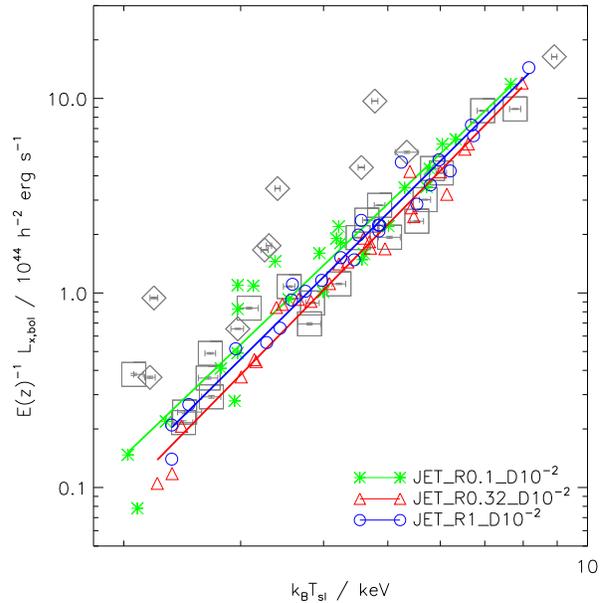}
\caption{The X-ray luminosity-temperature scaling relations predicted by our
  stochastic AGN feedback model with $f_{\rm rad}=0.1$ (asterisks), $0.32$
  (triangles) and $1$ (circles), keeping the other model parameter, $f_{\rm
    duty}$, fixed at $10^{-2}$. See the legend for model names. X-ray properties
  are computed within $r_{500}$. For comparative purposes, we also plot
  observational data for CC (diamonds) and NCC (squares) clusters in the REXCESS
  sample (PCA09).}
\label{fig:radvaryLT}
\end{figure}

Figure \ref{fig:radvaryLT} shows the \lxtsl\ relation for our full $25$-cluster
sample for the three different values of $f_{\rm rad}$ tested. Here also, we can
see that the three different models predict a very similar
\lxtsl\ relation. Note that the trend in \lxtsl\ with $f_{\rm rad}$ is not
monotonic: excessively large and excessively small values of $f_{\rm rad}$ will
both leave behind low-entropy core particles.  However, the variation is small
and each of the chosen values yields an adequate match to the observed relation for
NCC clusters, with $f_{\rm rad}=1$ providing the best match of the three.

\subsubsection{The effect of changing $f_{\rm duty}$}

\begin{figure}
\includegraphics[width=85mm]{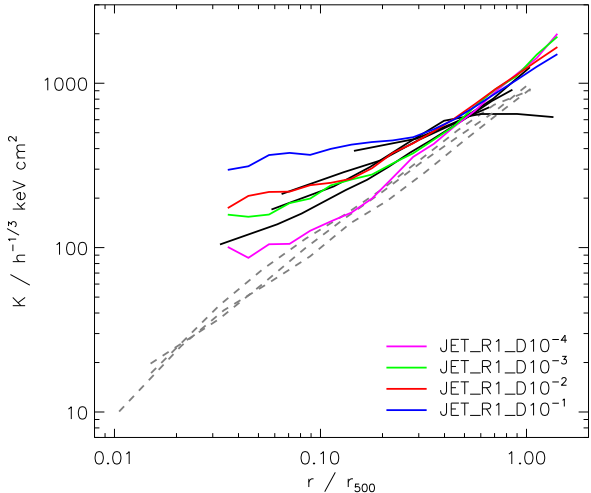}
\caption{Entropy profiles for cluster C1 resimulated with our stochastic AGN
  feedback model for several choices of the parameter $f_{\rm duty}$ (solid
  coloured lines; see legend for model details). The other model parameter,
  $f_{\rm rad}$, is set to unity. For comparison, we also show the profiles of
  observed CC (dashed grey lines) and NCC (solid black lines) clusters in the
  REXCESS sample (PAP10).}
\label{fig:fdutyvary}
\end{figure}

We now turn our attention to the effect of the parameter $f_{\rm duty}$. We have
done four runs, with $f_{\rm duty}=10^{-4}$, $10^{-3}$, $10^{-2}$ and $10^{-1}$,
respectively, keeping $f_{\rm rad}$ fixed at unity. The entropy profile of
cluster C1 in each case is displayed in Figure \ref{fig:fdutyvary}. It is
immediately clear that varying $f_{\rm duty}$ has a much larger effect on the
entropy of intracluster gas than $f_{\rm rad}$. As $f_{\rm duty}$ is increased
from $10^{-4}$ to $10^{-1}$, the slope of the entropy profile at radii
$r\lesssim 0.4r_{500}$ becomes progressively shallower. For $f_{\rm
  duty}=10^{-4}$ the slope is too steep compared to that of observed NCC cluster
profiles, whereas it is too flat for $f_{\rm duty}=10^{-1}$. The values $f_{\rm
  duty}=10^{-3}$ and $f_{\rm duty}=10^{-2}$ both give a good match to the
observational data for NCC clusters.


To explain the variation in cluster entropy profiles with $f_{\rm duty}$, we
have again examined what happens to particles that are heated by AGN feedback in
each of our runs. In the case where $f_{\rm duty}=10^{-4}$, the probability of a
particle being heated is low, but any particle that is heated receives a large
entropy boost since $P_{\rm heat}$ appears in the denominator of equation
(\ref{eq:stochasticdA}). The high entropy of heated particles causes them to
rise buoyantly to large cluster-centric radii, $r\gtrsim r_{500}$, leaving the
entropy profile in the core relatively undisturbed compared to a run with
SN feedback only. When $f_{\rm duty}$ is increased to $10^{-1}$, many more
particles in central regions are heated by AGN feedback since $P_{\rm heat}$ is
larger, and the entropy boost they are given is smaller. Accordingly, the
distance they move outwards from the core is less, resulting in a higher central
entropy and a flatter profile.

\begin{figure}
\includegraphics[width=85mm]{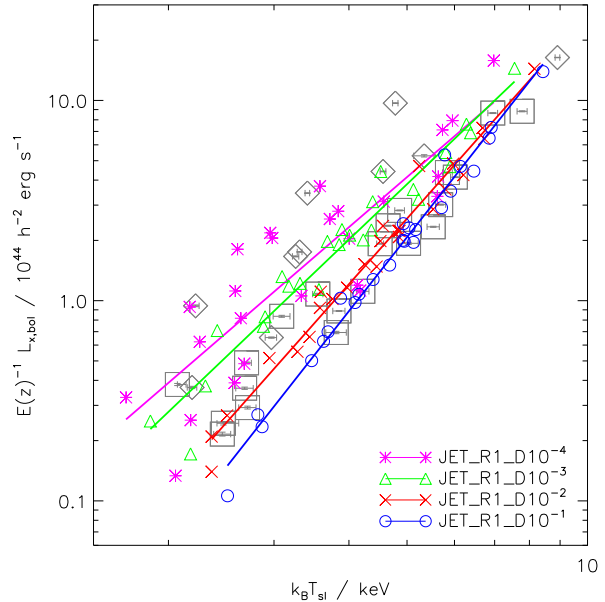}
\caption{The X-ray luminosity-temperature scaling relations predicted by our
  stochastic AGN feedback model with $f_{\rm duty}=10^{-4}$ (asterisks),
  $10^{-3}$ (triangles), $10^{-2}$ (crosses) and $10^{-1}$ (circles), keeping
  the other model parameter, $f_{\rm rad}$, set to unity. See the legend for
  model names. X-ray properties are computed within $r_{500}$. Observed CC
  (diamonds) and NCC (squares) clusters from the REXCESS (PCA09) are also
  displayed.}
\label{fig:fdutyvaryLT}
\end{figure}

The large impact of $f_{\rm duty}$ on cluster entropy profiles is reflected in
the \lxtsl\ scaling relation. This is demonstrated in Figure
\ref{fig:fdutyvaryLT} where we show the \lxtsl\ relation for our full cluster
sample predicted by each of our four models. For $f_{\rm duty}=10^{-4}$, the
low central entropy causes an enhanced X-ray luminosity, so all of our simulated
clusters lie well above the mean observed relation for NCC clusters. In fact,
the predicted relation in this case resembles the observed relation for CC
clusters, although this is artificial since we have not included cooling
processes. As $f_{\rm duty}$ is increased, the normalisation of the
\lxtsl\ relation decreases and the slope becomes steeper. A good match to the
observed NCC cluster \lxtsl\ relation is obtained when $f_{\rm
  duty}=10^{-2}$. For larger values of $f_{\rm duty}$, the relation is too
steep, so that low-temperature systems have too low a luminosity for their
mass. This is because the AGN heating has raised the core entropy in these
systems to an excessive level.

\subsubsection{Identifying optimal parameter values using observations}

The next issue to address is whether observational data can help us to constrain
the two free parameters of our stochastic AGN feedback model. For this analysis,
we have resimulated all $25$ clusters in our sample (C1--C25) using
different combinations of the pair of parameters $(f_{\rm rad},f_{\rm
  duty})$. The values adopted are $f_{\rm rad}=0.1$, $0.32$ and $1$, and $f_{\rm
  duty}=10^{-4}$, $10^{-3}$, $10^{-2}$ and $10^{-1}$, giving a grid of $12$
models in total.

We assess the suitability of each model by testing how well it reproduces the
observed scaling of three fundamental ICM observables with spectroscopic
temperature: (i) the entropy profile normalisation, which we take to be the
entropy at $r_{1000}$ (typically about $0.7r_{500}$), (ii) the entropy profile
shape, defined as the ratio of the entropy at $r_{1000}$ to the entropy measured
at $0.1r_{200}$ (i.e. the central entropy), and (iii) the X-ray
luminosity. Again, the source of the observational data is the REXCESS. Note
that we are only aiming to match the observed scaling relations for NCC clusters
since cooling is not included in our simulations at this stage.

In the following analysis, we neglect any clusters in our simulated sample that
have large amounts of substructure. To differentiate between dynamically relaxed
and disturbed systems, we use the substructure statistic
\begin{equation}
S=\frac{|\mathbf{x}_{\rm com}-\mathbf{x}_{\rm c}|}{r_{500}},
\end{equation}
where $\mathbf{x}_{\rm c}$ is the location of the dark matter
potential minimum and $\mathbf{x}_{\rm com}$ is the centre of mass of
the cluster within $r_{500}$. Following \citet{KDA07}, we say that a
cluster is disturbed if $S>0.1$, and relaxed otherwise.

For each scaling relation, the criterion we use to test how well our models
reproduce the mean observed relation is the $\chi^2$ statistic
\begin{eqnarray}
\label{eq:chisq}
\chi^2&=&\frac{1}{\sigma_{\rm int}^2}\sum_{i=1}^{N_{\rm c}^{\rm
    sim}}\left\{\log_{10}{\left[E(z)^nY_i^{\rm
      sim}\right]}\right.\\ &&\hspace*{1.5cm}\left.-\alpha\log_{10}{\left(\frac{T_{{\rm
        sl},i}^{\rm sim}}{5\ {\rm
      keV}}\right)-\log_{10}{C_0}}\right\}^2,\nonumber
\end{eqnarray}
where $N_{\rm c}^{\rm sim}$ is the number of (relaxed) simulated clusters and
$Y=K(r_{1000})$, $K(r_{1000})/K(0.1r_{200})$ or $\lx$, depending on
the relation being considered. The quantities $C_0$ and $\alpha$ are
the normalisation and slope of a power-law fit to the corresponding
observed relation,
\begin{equation}
\label{eq:genscalerel}
E(z)^n Y^{\rm obs}=C_0\left(\frac{T_{\rm X}^{\rm obs}}{5\ {\rm keV}}\right)^{\alpha},
\end{equation}
obtained by using the BCES orthogonal linear regression method
\citep{AKB96} in log-log space, taking into account the errors in both
$T_{\rm X}^{\rm obs}$ any $Y^{\rm obs}$. The normalisation $C_0$ has
units of $h^{-1/3}\,{\rm keV}$ cm$^2$ and $10^{44}h^{-2}\ {\rm
  erg\ s}^{-1}$ for $Y=K(r_{1000})$ and $\lx$, respectively, and is
dimensionless for $Y=K(r_{1000})/K(0.1r_{200})$. The factor $E(z)^n$
is included to remove the predicted self-similar evolution, where the
index $n=4/3$, $0$ and $-1$ for the $K(r_{1000})$-$T_{\rm X}$,
$K(r_{1000})/K(0.1r_{200})$-$T_{\rm X}$ and $L_{\rm X}$-$T_{\rm X}$
relations, respectively. 

The scatter expected from statistical uncertainties, $\sigma_{\rm stat}$, is
\begin{equation}
\label{eq:sigstat}
\sigma_{\rm stat}^2=\frac{1}{(1/N_{\rm c}^{\rm obs})\sum_{i=1}^{N_{\rm c}^{\rm obs}}1/\sigma_i^2},
\end{equation} 
where
\begin{equation}
\sigma_i^2=(\sigma_{{\rm Y},i}^{\rm obs})^2+\alpha^2(\sigma_{T_{\rm X},i}^{\rm obs})^2,
\end{equation}
and $\sigma_{T_{\rm X},i}^{\rm obs}$ and $\sigma_{{\rm Y},i}^{\rm
  obs}$ are the errors in $T_{{\rm X},i}^{\rm obs}$ any $Y_i^{\rm
  obs}$, respectively. 

We estimate the raw scatter, $\sigma_{\rm raw}$, using error-weighted
distances to the regression line:
\begin{eqnarray}
\label{eq:sigraw}
\sigma_{\rm raw}^2&=&\frac{1}{N_{\rm c}^{\rm obs}-2}\sum_{i=1}^{N_{\rm c}^{\rm
    obs}}w_i\left\{\log_{10}{\left[E(z)^nY_i^{\rm obs}\right]}\right.\\
&&\hspace*{1.5cm}\left.-\alpha\log_{10}{\left(\frac{T_{{\rm X},i}^{\rm
      obs}}{5\ {\rm keV}}\right)-\log_{10}{C_0}}\right\}^2,\nonumber
\end{eqnarray} 
where $w_i=\sigma_{\rm stat}^2/\sigma_i^2$ and $N_{\rm c}^{\rm obs}$ is
the number of observed NCC clusters in the sample.

Finally, the intrinsic scatter, $\sigma_{\rm int}$, about each
observed mean relation is estimated as
\begin{equation}
\label{eq:sigint}
\sigma_{\rm int}^2=\sigma_{\rm raw}^2-\sigma_{\rm stat}^2.
\end{equation}

We examine how well our models reproduce the observed scatter about
each mean relation, by using the Kolmogorov-Smirnov (K-S) test to
determine if the residuals for the observed and simulated samples are
drawn from the same distribution in each case.

Tables \ref{tab:LTchisq} and \ref{tab:LTKS} show the probabilities
($p$-values) for each of our $12$ models obtained from the $\chi^2$
(equation \ref{eq:chisq}) and K-S tests, respectively, for the case of
the \lxtx\ relation. The models we deem to be acceptable are
highlighted in bold. It is evident that all models with $f_{\rm
  duty}\leq 10^{-3}$ are ruled out, at least for this particular
relation, and only three models provide an acceptable match to both
the mean observed relation and the associated scatter.

\begin{table}
\caption{$\chi^2$ test probability values for the $\lx$-$T_{\rm sl}$ scaling relation.}
\label{tab:LTchisq}
\begin{tabular}{@{}lccc}
\hline
\backslashbox{$f_{\rm duty}$}{$f_{\rm rad}$} & 0.1 & 0.32 & 1.0 \\
\hline
$10^{-4}$ & 0.00 & 0.00 & 0.00 \\
$10^{-3}$ & 0.00 & 0.00 & 0.00 \\
$10^{-2}$ & 0.00 & {\bf 0.20} & {\bf 0.96} \\
$10^{-1}$ & {\bf 0.69} & 0.00 & 0.01 \\
\hline
\end{tabular}
\end{table}

\begin{table}
\caption{K-S test probability values for the $\lx$-$T_{\rm sl}$ scaling relation.}
\label{tab:LTKS}
\begin{tabular}{@{}lccc}
\hline
\backslashbox{$f_{\rm duty}$}{$f_{\rm rad}$} & 0.1 & 0.32 & 1.0 \\
\hline
$10^{-4}$ & 0.00 & 0.01 & 0.00 \\
$10^{-3}$ & 0.00 & 0.02 & 0.00 \\
$10^{-2}$ & {\bf 0.41} & {\bf 0.12} & {\bf 0.24} \\
$10^{-1}$ & {\bf 0.73} & 0.00 & 0.02 \\
\hline
\end{tabular}
\end{table}

For the sake of brevity, we do not present the corresponding tables
for the $K(r_{1000})$-$T_{\rm X}$ and
$K(r_{1000})/K(0.1r_{200})$-$T_{\rm X}$ relations. We simply note that
the $\chi^2$ test for the $K(r_{1000})$-$T_{\rm X}$ relation rules out
all models expect the two with $(f_{\rm rad},f_{\rm
  duty})=(1,10^{-3})$ and $(1,10^{-2})$, whereas the K-S test rules
out all models with $f_{\rm duty}=10^{-4}$ or $f_{\rm
  duty}=10^{-1}$. In the case of the
$K(r_{1000})/K(0.1r_{200})$-$T_{\rm X}$ relation, both the $\chi^2$
and K-S tests rule out all models with $f_{\rm duty}=10^{-1}$, tending
to favour models that populate the upper-right corner of the table.

We have combined the results of all $6$ tests ($2$ for each of the $3$
scaling relations) using Fisher's method for combining $p$-values. The
overall probabilities for our $12$ models are summarised in Table
\ref{tab:totchisq}. It is evident that only one model is now
acceptable, the model with $(f_{\rm rad},f_{\rm
  duty})=(1,10^{-2})$. Therefore, we choose our fiducial AGN feedback
scheme to be the stochastic model JET\_R1\_D$10^{-2}$.

\begin{table}
\caption{Combined probability values for all three scaling relations.}
\label{tab:totchisq}
\begin{tabular}{@{}lccc}
\hline
\backslashbox{$f_{\rm duty}$}{$f_{\rm rad}$} & 0.1 & 0.32 & 1.0 \\
\hline
$10^{-4}$ & 0.00 & 0.00 & 0.00 \\
$10^{-3}$ & 0.00 & 0.00 & 0.00 \\
$10^{-2}$ & 0.00 & 0.00 & {\bf 0.34} \\
$10^{-1}$ & 0.00 & 0.00 & 0.00 \\
\hline
\end{tabular}
\end{table}

We emphasise that the purpose of this section has not been to conduct
a rigorous statistical analysis, but merely to provide us with an
indication of the region of our model parameter space that is favoured
by the observational data. There are several possible caveats to our
analysis. For example, we are assuming Gaussian errors and that each
test is strictly independent, both of which may not be the case.

\subsection{Metallicity profiles}
\label{sec:agnmet}

Powerful AGN outflows are an obvious candidate for transporting metal-rich
material away from the central regions of halos. Indeed, there is observational
evidence that enriched gas is entrained by bubbles inflated by central AGN and
removed from cluster cores (e.g. \citealt{FNH05,MWS10,KMC11}). Cosmological
hydrodynamical simulations have also demonstrated that AGN are important for the
metal enrichment of intracluster gas (e.g. \citealt{SSD07,MSD07,FBT10,WST11}).

Recall from Section \ref{sec:snmet} that SN feedback alone has a negligible
impact on the distribution of metals in the ICM, regardless of the way in which
the available energy is injected into the diffuse gas. We now investigate
whether the inclusion of extra energy input from AGN affects ICM metallicity
profiles. To do this, we have resimulated cluster C1 with our fiducial AGN
feedback model (JET\_R1\_D$10^{-2}$), varying the parameter $f_{\rm Z,rad}$ in
our metal enrichment scheme (recall that this parameter sets the radius of the
spherical region about a galaxy within which metals are injected).

\begin{figure}
\includegraphics[width=85mm]{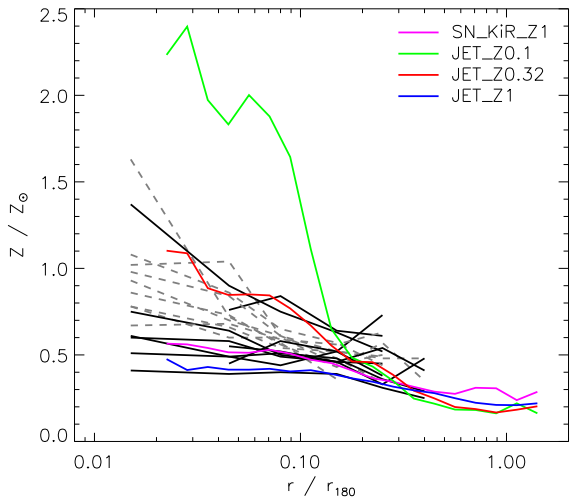}
\caption{Emission-weighted metallicity profiles for cluster C1 resimulated with
  our fiducial stochastic AGN feedback model, but with different metal
  enrichment schemes (solid coloured lines; see legend for model details). The
  profile obtained from a run with supernova feedback only is shown for
  comparison, as well as the profiles of observed CC (dashed grey lines) and NCC
  (solid black lines) clusters from the sample of MAT11.}
\label{fig:agnZprof}
\end{figure}

Figure \ref{fig:agnZprof} compares the resulting emission-weighted
metallicity profiles with those of observed clusters from the sample
of MAT11. As before, the best match to the gradient of the profiles of observed NCC
clusters arises when metals are distributed uniformly throughout the
entire halo ($f_{\rm Z,rad}=1$). As $f_{\rm Z,rad}$ is decreased, we
see the development of a sharp central abundance peak that is in
conflict with the observational data. Recall that a distributed metal
enrichment model is justified since the \SAM\ underpinning our
simulations predicts that almost all metals in the ICM are accreted,
rather than being produced by BCGs.

Comparing the metallicity profile predicted by model JET\_Z1 to that obtained
from the SN\_KiR\_Z1 run, we can see that addition of AGN feedback has indeed
displaced some metals from central cluster regions, leading to a flatter
profile, but the effect is small. This is because only a small fraction of the
particles in the core are heated by AGN in our model, and they receive a
sufficiently large entropy boost to escape the central regions of clusters,
leaving the majority of metal-rich material behind in the core.

\subsection{Summary}

We have developed a new AGN feedback model where heat energy is
injected into intracluster gas in a stochastic manner. Our model is
physically-motivated and has just two free parameters: $f_{\rm rad}$,
which sets the radius of the spherical region within which energy is
injected, and $f_{\rm duty}$, which is the fraction of particles in
this region that are heated per AGN duty cycle.

We have found that ICM entropy profiles and the \lxtsl\ scaling relation
are fairly insensitive to variations in $f_{\rm rad}$, but depend
strongly on $f_{\rm duty}$. For small values of $f_{\rm duty}$, the
resulting entropy profiles are close to those obtained from a run with
SN feedback only, implying that AGN heating has little effect
in this case. This is because only a few particles are heated, and
they receive large entropy boosts, which causes them to rise buoyantly
to large distances from the cluster centre, leaving the majority of
low-entropy gas behind.

As $f_{\rm duty}$ is increased, the probability of a particle being
heated also increases: hence more particles in the core are heated,
they are given a smaller entropy injection, and so they do not escape
central cluster regions. As the heated gas expands, the gas density
drops, causing the gradient of the resulting entropy profiles to
become shallower.

We have used three observed scaling relations to identify an optimal
choice for the two free parameters in our model: $(f_{\rm rad},f_{\rm
  duty})=(1,10^{-2})$. Setting $f_{\rm duty}=10^{-2}$ roughly
corresponds to a jet opening angle of $16^{\circ}$. With these
parameter choices, our model, named JET\_R1\_D$10^{-2}$, can explain
the observed entropy profiles and \lxt\ relations for NCC clusters.

Using JET\_R1\_D$10^{-2}$ as our fiducial AGN feedback model, we have
demonstrated that AGN heating has little impact on the distribution of
metals in the ICM by comparing to a model with kinetic supernova
feedback only. When the metals produced by stars in galaxies are
distributed uniformly throughout the entire host halo (model JET\_Z1),
the resulting abundance gradients provide a good match to those
observed in NCC clusters.

For the remainder of this paper we thus adopt model JET\_Z1 as our
fiducial model for star formation, metal production, black hole growth
and associated stellar and AGN feedback.

\section{Comparison with observations}
\label{sec:obscomp}

In this section we conduct a detailed assessment of how well our
fiducial model (JET\_Z1) can reproduce key observed thermal and chemical
properties of intracluster gas.

\subsection{Thermal properties of the ICM}

\begin{figure}
\includegraphics[width=85mm]{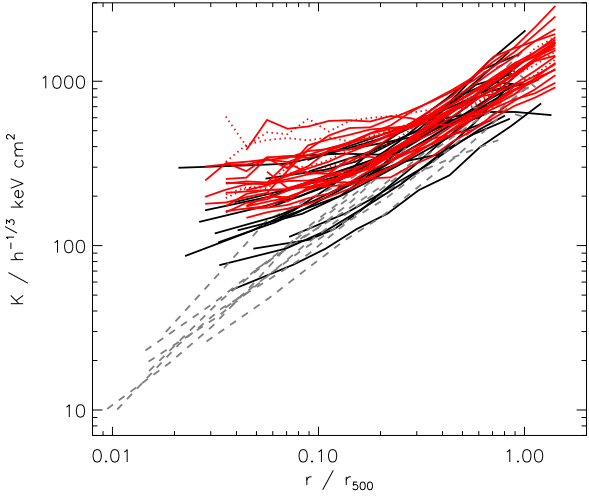}
\caption{Entropy profiles for $25$ clusters resimulated with our fiducial
  stochastic AGN feedback model. The profiles of relaxed (disturbed) systems are
  shown by solid (dotted) red lines. Observed profiles of CC (dashed grey lines)
  and NCC (solid black lines) clusters in the REXCESS sample (PAP10) are also
  displayed for comparative purposes. The observed clusters span the same mass
  range as our simulated ones.}
\label{fig:bestKprof}
\end{figure}

Figure \ref{fig:bestKprof} compares the predicted entropy profiles of
all $25$ clusters (C1--C25) in our sample with the profiles of observed
systems in the same mass range. The profiles of relaxed
(disturbed) simulated clusters are shown as solid (dotted) red lines.
 
First impressions are that our fiducial model generates clusters whose
entropy profiles agree well with those of observed NCC systems, both
in terms of normalisation and gradient. The central entropy is too
high in three of our objects, but two of these are classified as
disturbed systems. To assess our model more quantitatively, we
now examine how the entropy profile normalisation and shape scale with
system temperature.

\begin{figure}
\includegraphics[width=85mm]{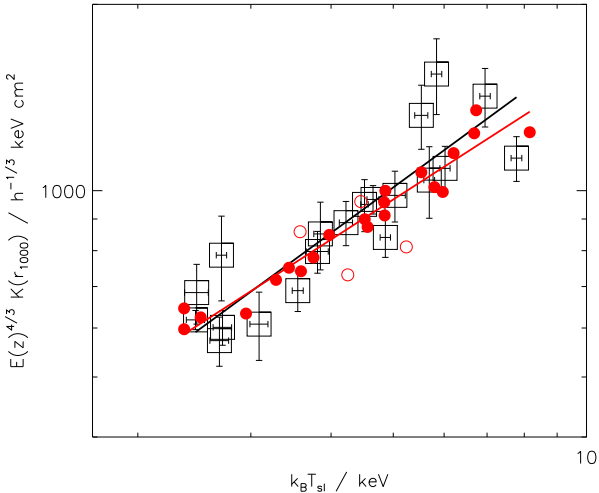}
\caption{The scaling of entropy profile normalisation, $K(r_{1000})$, with
  temperature predicted by our fiducial stochastic AGN feedback model. We
  compute the spectroscopic-like temperature within $r_{500}$. Filled (open)
  circles correspond to relaxed (disturbed) systems in our simulated cluster
  sample, and the solid red line is the best-fit relation considering relaxed
  systems only. The observed relation for NCC clusters in the REXCESS sample
  (PAP10) is also shown by open squares and a solid black line.}
\label{fig:bestKnorm}
\end{figure}

Figure \ref{fig:bestKnorm} shows the entropy profile normalisation (defined as
the entropy at $r_{1000}$) as a function of spectroscopic-like
temperature. Filled (open) circles correspond to relaxed (disturbed) objects
(both here and in all subsequent figures). Observational data for NCC clusters
in the REXCESS are also shown. The parameters of the accompanying predicted and
observed best-fit relations are summarised in Table \ref{tab:scalerel}. Note
that we only consider relaxed systems in our sample when performing the fit and,
as before, we adopt the BCES orthogonal fitting method. It is evident that the
\knormtsl\ relation predicted by our model is a good match to the observed
relation: the slope and normalisation are both within $1\sigma$ of the observed
values, and the scatter about the mean relation is comparable.

\begin{table*}
\begin{minipage}{126mm}
\caption{Best-fit parameters (with $1\sigma$ errors) for $z=0$ scaling relations
  obtained from our full 25-cluster simulated sample, and from the REXCESS
  observations of PAP10. Note that we only consider relaxed clusters in our
  sample when deriving predicted relations. All fits were performed using the
  BCES orthogonal regression method.}
\label{tab:scalerel}
\begin{tabular}{@{}lcccccc}
\hline
Relation & & Predicted & & & Observed & \\
\hline
& $C_0$ & $\alpha$ & $\sigma_{\rm int}$ & $C_0$ & $\alpha$ & $\sigma_{\rm int}$ \\
\hline
$K(r_{1000})$-$T_{\rm sl}$ & $968\pm 15$ & $0.667\pm 0.044$ & $0.025$ & $1013\pm
41$ & $0.76\pm 0.11$ & $0.032$ \\ 
$K(r_{1000})/K(0.1r_{200})$-$T_{\rm sl}$ & $2.81\pm 0.16$ & $0.57\pm 0.33$ &
$0.12$ & $3.31\pm 0.35$ & $0.49\pm 0.45$ & $0.12$\\ 
$L_{\rm X}$-$T_{\rm sl}$ & $2.530\pm 0.080$ & $3.41\pm 0.11$ & $0.071$ &
$2.43\pm 0.13$ & $3.22\pm 0.12$ & $0.098$ \\ 
$Z(0.25r_{180})/Z(0.045r_{180})$-$T_{\rm sl}$ & $0.705\pm 0.019$ & $-0.071\pm
0.086$ & $0.055$ & $0.824\pm 0.067$ & $-0.11\pm 0.14$ & $0.029$ \\ 
\hline
\end{tabular}

\medskip
$C_0$ and $\alpha$ are the best-fitting normalisation and slope of the
relations, respectively (see equation \ref{eq:genscalerel}), and $\sigma_{\rm
  int}$ is the intrinsic scatter about the mean relation (equation
\ref{eq:sigint}).
\end{minipage}
\end{table*}

\begin{figure}
\includegraphics[width=85mm]{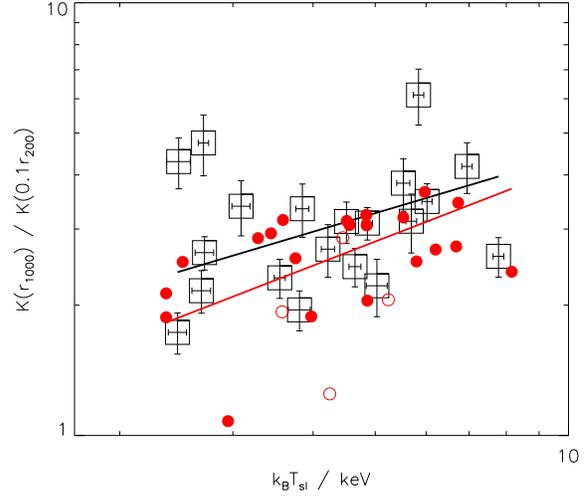}
\caption{The scaling of entropy profile shape, $K(r_{1000})/K(0.1r_{200})$, with
  temperature predicted by our fiducial stochastic AGN feedback model. The
  spectroscopic-like temperature is computed within $r_{500}$.  Relaxed
  (disturbed) systems in our simulated cluster sample are shown as filled (open)
  circles. The solid red line is the best-fit relation obtained using our
  relaxed clusters only. For comparison, we also display the observed relation
  for NCC clusters from the REXCESS (PAP10; open squares and black
    line).}
\label{fig:bestKrat}
\end{figure}

In Figure \ref{fig:bestKrat} we display the variation of the ratio
$K(r_{1000})/K(0.1r_{200})$ (a measure of the entropy profile shape) with
temperature. The slope of our predicted relation is consistent with that of the
observed relation and the scatter is identical to the observed value; see Table
\ref{tab:scalerel}.  However, the normalisation is slightly lower. There are two
reasons for this offset. First, one of our relaxed clusters has an anomalously
low value of $K(r_{1000})/K(0.1r_{200})$, which lowers the normalisation of the
predicted \kshapetsl\ relation. This objects correspond to the relaxed system
with an excessive central entropy in Figure \ref{fig:bestKprof}. Second, three
of the observed clusters lie considerably above any of our simulated clusters on
the \kshapetsl\ plane. This acts to raise the normalisation of the observed
relation relative to the predicted one. Although classified as NCC systems in
REXCESS, these objects actually have a low central entropy, reminiscent of CC
clusters; see Figure \ref{fig:bestKprof}. Without these outliers, there is good
overall agreement between the predicted and observed \kshapetsl\ relations.

\begin{figure}
\includegraphics[width=85mm]{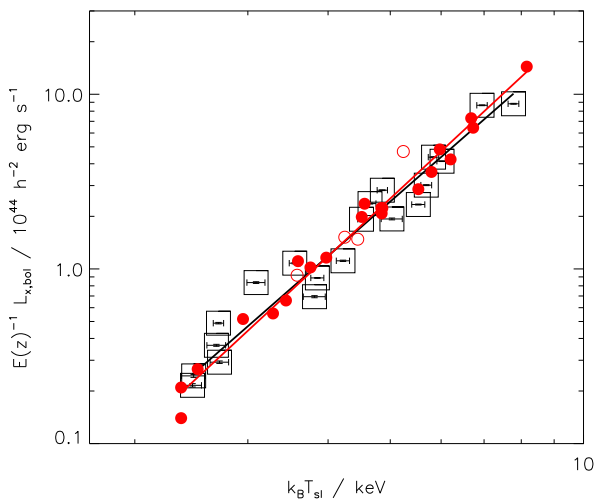}
\caption{The X-ray luminosity-temperature scaling relation predicted by our
  fiducial stochastic AGN feedback model. X-ray properties are computed within
  $r_{500}$. Filled (open) circles correspond to relaxed (disturbed) systems in
  our simulated cluster sample, and the solid red line is the best-fit relation
  for relaxed objects only. For comparative purposes, we also plot the observed
  relation for NCC clusters in the REXCESS sample (PCA09; squares and solid
  black line).}
\label{fig:bestLT}
\end{figure}

Finally, we contrast our predicted \lxtsl\ scaling relation with the REXCESS NCC
cluster relation in Figure \ref{fig:bestLT}. We predict slightly less scatter
about the mean relation than observed, but we recover the normalisation and
slope of the observed relation to within $1\sigma$, as summarised in Table
\ref{tab:scalerel}. We conclude that our fiducial model yields an
\lxtsl\ relation that is a good match to the observed relation for NCC clusters.

\subsection{Chemical properties of the ICM}

\begin{figure}
\includegraphics[width=85mm]{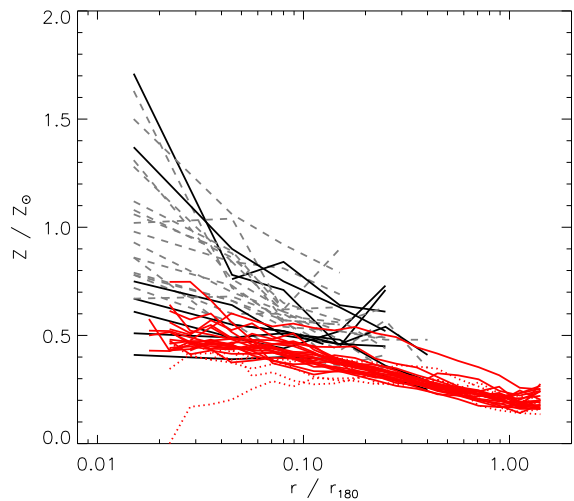}
\caption{Emission-weighted metallicity profiles for $25$ clusters resimulated
  with our fiducial stochastic AGN feedback model. The profiles of relaxed
  (disturbed) systems are shown by solid (dotted) red lines. For comparison, we
  also show observed profiles of CC (dashed grey lines) and NCC (solid black
  lines) clusters in the REXCESS sample (PAP10). We only show observed clusters
  with a mass in the same range as our simulated objects.}
\label{fig:bestZprof}
\end{figure}

The predicted emission-weighted metallicity profiles of all $25$ clusters in our
sample are displayed in Figure \ref{fig:bestZprof}, along with Fe abundance
profiles of observed CC and NCC clusters from the sample of MAT11. To ensure a
fair comparison, we only plot the profiles of observed clusters that lie in the
mass range spanned by our simulated sample. Again, solid (dotted) red
lines correspond to relaxed (disturbed) systems.  We remind the reader of the
limitations of our metallicity model: we assume only prompt enrichment and
cannot discriminate between ejecta from core collapse and Type 1a SN.
Nevertheless, it is evident that the profiles of our simulated clusters are in
reasonable agreement with those of observed NCC clusters, both in terms of
normalisation and slope, with the exception of the few observed NCC objects that
have a sharp central abundance peak, which could be CC remnants
\citep{LRM10,ROM10}. To demonstrate this more rigorously, we now investigate how
the metallicity profile shape scales with temperature.

The measure of metallicity profile shape we adopt is the ratio of the
metallicity at a radius of $0.25r_{180}$ to that at a radius of
$0.045r_{180}$. We chose these particular radii since nearly all of the clusters
in the sample of MAT11 have a metallicity profile defined over this radial
range, thereby maximising the number of observed clusters we can compare our
predictions to.

\begin{figure}
\includegraphics[width=85mm]{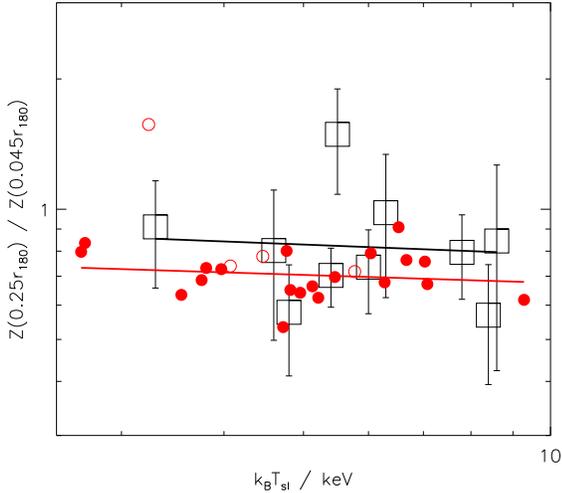}
\caption{The scaling of emission-weighted metallicity profile shape,
  $Z(0.25r_{180})/Z(0.045r_{180})$, with temperature predicted by our fiducial
  stochastic AGN feedback model. The filled (open) circles correspond to relaxed
  (disturbed) systems in our simulated cluster sample. The solid red line is the
  best-fit relation obtained when considering just the relaxed systems in our
  sample. The observed relation for NCC clusters in the sample of MAT11 is also
  displayed (squares and solid black line).}
\label{fig:bestZrat}
\end{figure}

Figure \ref{fig:bestZrat} shows the predicted scaling of the ratio
$Z(0.25r_{180})/Z(0.045r_{180})$ with spectroscopic-like temperature. The
corresponding observed relation for NCC clusters is also shown for comparison,
and the parameters of both best-fit relations are presented in Table
\ref{tab:scalerel}. We recover the observed gradient to within $1\sigma$, but
the predicted normalisation is lower than observed, and the scatter is larger.
However, the observational errors are large and there is one observed cluster
that lies considerably above all the others, which will act to increase the
normalisation of the observed relation relative to that of ours.

\section{Including radiative cooling: A first attempt}
\label{sec:cooling}

None of the simulations presented in this paper thus far incorporate cooling
processes. In this section, we make a first attempt to extend our hybrid
feedback scheme by allowing gas to cool radiatively. Our aim is to formulate a
feedback model which can produce \emph{both} CC and NCC clusters, whilst
avoiding catastrophic over-cooling of gas in central cluster regions. We
emphasise that this work is exploratory, intended merely to demonstrate that
such a model is possible with our approach.

The addition of gas cooling is likely to lead to differences between the
predictions of kinetic and thermal feedback schemes that have not been apparent
in our previous non-radiative runs. However, it is not our intention here to
conduct an exhaustive comparison of different feedback models when cooling
processes are included; we save this for future work.

We implement AGN feedback using the $2$-parameter stochastic heating model
developed in the Section~\ref{sec:newagn} (however, as we shall see, the optimal
parameter choices change with the addition of cooling), and we adopt our
fiducial model for SN feedback (gas particles neighbouring a galaxy are given a
kick in a random direction with a speed of $600$ km s$^{-1}$) and metal
enrichment (metals produced by stars in galaxies are uniformly distributed
throughout the entire host halo).

Metal-dependent radiative cooling is included in our simulations as follows. For
each gas particle, we know its (smoothed) metallicity, $Z_{{\rm sm},i}$ (see
equation \ref{eq:Zsm}), and we can compute its temperature from its entropy,
$A_i$, and density, $\rho_i$. With this information we then calculate the
cooling rate using the cooling function of \citet{SUD93}, and reduce the entropy
of the gas particle accordingly.

\begin{figure}
\includegraphics[width=85mm]{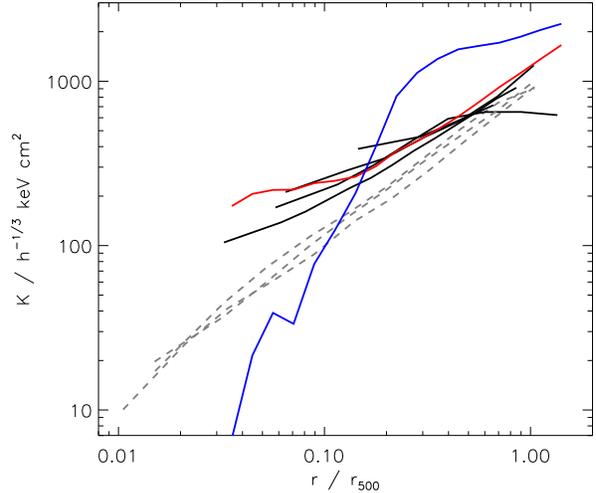}
\caption{Entropy profiles for cluster C1 resimulated with our fiducial
  stochastic AGN feedback model without cooling (solid red line), and with
  metal-dependent radiative cooling (solid blue line). Observed profiles of
  similar-mass CC (dashed grey lines) and NCC (solid black lines) clusters in
  the REXCESS sample (PAP10) are also displayed for comparative purposes.}
\label{fig:overcool}
\end{figure}

Figure \ref{fig:overcool} compares the entropy profile of cluster C1 obtained
from runs with our fiducial feedback model (JET\_Z1) with and without
cooling. It is apparent that the heating from SN and AGN has not been sufficient
to prevent over-cooling in central cluster regions: there is a sharp drop in gas
temperature at $r\lesssim 0.3r_{500}$, leading to a steep decline in the entropy
profile, and there is an entropy increase at larger radii due to hotter,
lower-density gas flowing inwards from cluster outskirts to maintain pressure
support in the core.

\emph{A priori}, there is no reason to expect that the amount of SN and AGN
heating provided by the underlying SA model would be sufficient to precisely
balance radiative cooling in the simulation. This is because \lgalaxies\ employs
a simple cooling recipe based on the assumption that haloes have a
spherically-symmetric isothermal gas distribution, which is typically not the
case in hydrodynamical simulations, so the predicted cooling rate of gas in
haloes will be different to that in the simulation. Since the amount of gas that
can cool to form stars and accrete onto central black holes governs the level of
subsequent feedback, such differences in gas cooling rates imply that it is
unlikely a self-regulating feedback loop would be established in the simulation.

To address this problem, we have developed an \emph{ad hoc} extension of our
stochastic AGN feedback model where we inject extra energy into cluster cores as
a crude representation of additional AGN heating that would have arisen from
enhanced black hole accretion due to more efficient cooling. Such a scheme is
justifiable, provided the extra energy input required to balance radiative
cooling is a small fraction of that originally available from the SA model.

The details of our model are as follows. At each SA model output, we identify
all gas particles in the simulation residing in the central regions of haloes
($r<0.1r_{200}$). At each subsequent timestep, we test if any of these particles
have cooled below a threshold temperature of $3\times 10^4$ K; if they have, we
raise their temperature to some multiple, $f_{\rm temp}$, of the virial
temperature, $T_{200}$ (equation \ref{eq:T200}), of their host halo at the
previous output time. We continue in this fashion until the next model output is
reached, at which point the list of particles contained in halo cores is reset
and the process is repeated.

In what follows we keep the radius of energy injection in our AGN feedback model
fixed at unity ($f_{\rm rad}=1$). We then have two free parameters: the fraction
of particles heated per AGN duty cycle, $f_{\rm duty}$, and $f_{\rm temp}$,
which controls the temperature cold particles in cluster cores are heated to. We
now explore the effect of varying these parameters on the entropy distribution
in clusters. All of our models are summarised in Table \ref{tab:coolagnmodels},
where we have given them the label ZCOOL to emphasise that they include
metal-dependent radiative cooling.

\begin{table*}
\begin{minipage}{126mm}
\caption{Stochastic AGN feedback models with metal-dependent radiative cooling
  and a prescription for additional heating of cold gas in cluster cores. In
  each case, supernova feedback is implemented using a kinetic model where
  particles neighbouring a galaxy are given a kick in a random direction with
  velocity $600$ km s$^{-1}$. Energy and metals are both injected within a
  radius of $r_{200}$ ($f_{\rm rad}=1$ and $f_{\rm Z,rad}=1$,
  respectively). Unless otherwise stated, the fraction of particles heated per
  AGN duty cycle is $f_{\rm duty}=10^{-1}$, and cold particles in cluster cores
  are heated to a temperature of $f_{\rm temp}=2.5$ times the halo virial
  temperature.}
\label{tab:coolagnmodels}
\begin{tabular}{@{}llll}
\hline
Model name & Type & Energy injection method & Comments \\
\hline
ZCOOL\_D$f_{\rm duty}$ & Stochastic & Fixed entropy &  $f_{\rm duty}=10^{-4}$, $10^{-3}$, $10^{-2}$, $10^{-1}$ \\
\hline
ZCOOL\_T$f_{\rm temp}$ & Stochastic & Fixed entropy &  $f_{\rm temp}=1,2.5,4.5,20$\\
\hline
\end{tabular}
\end{minipage}
\end{table*}

\subsection{The effect of changing $f_{\rm duty}$}

\begin{figure}
\includegraphics[width=85mm]{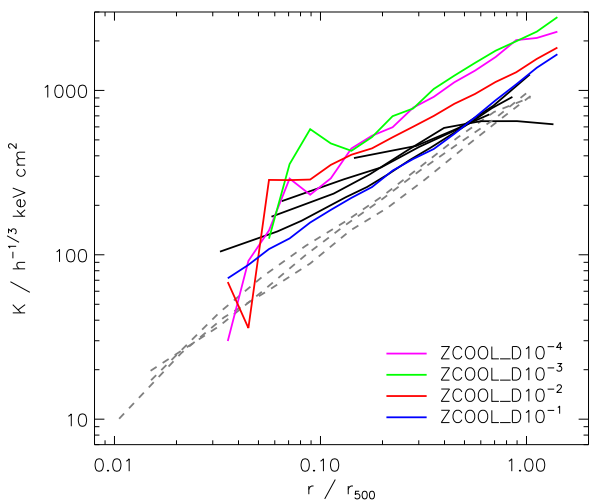}
\caption{Entropy profiles for cluster C1 resimulated with our
  stochastic AGN feedback model, including metal-dependent radiative
  cooling, for various choices of the parameter $f_{\rm duty}$
  (coloured lines; see legend for model details -- note that the
  exessive cooling for $f_{\rm duty}=10^{-4}$ has led to
  reclassification of the cluster as having high substructure). We
  also assume that any cold gas remaining in cluster cores is heated
  to $2.5$ times the halo virial temperature. For comparison, we also
  show the profiles of observed CC (dashed grey lines) and NCC (solid
  black lines) clusters in the REXCESS sample (PAP10).}
\label{fig:coolfduty}
\end{figure}

Figure \ref{fig:coolfduty} illustrates how the entropy profile of
cluster C1 is affected by varying $f_{\rm duty}$, keeping $f_{\rm
  temp}$ fixed at $2.5$. With the exception of $f_{\rm duty}=10^{-1}$,
all values of $f_{\rm duty}$ produce entropy profiles that exhibit
signs of over-cooling. This is because only a small fraction of core
particles are heated by AGN feedback in these cases, so gas can cool
efficiently in the cluster core, even with the injection of additional
energy. When we increase $f_{\rm duty}$ to $10^{-1}$, so that a larger
fraction of core particles are heated by AGN, we obtain a CC-like
entropy profile and it appears that radiative cooling has been
balanced. Therefore, we change our fiducial value of $f_{\rm duty}$
from $10^{-2}$ to $10^{-1}$, which corresponds to a larger jet opening
angle of about $52^{\circ}$.

\subsection{The effect of changing $f_{\rm temp}$}

\begin{figure}
\includegraphics[width=85mm]{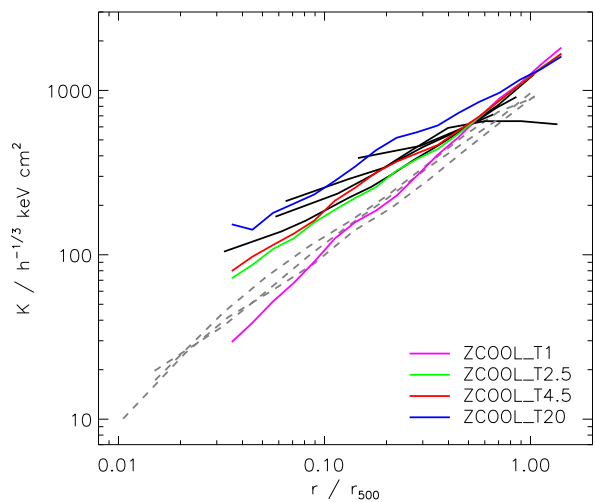}
\caption{Entropy profiles for cluster C1 resimulated with our
  stochastic AGN feedback model, including metal-dependent radiative
  cooling, and assuming that cold gas in the central regions of
  clusters is heated to some multiple, $f_{\rm temp}$, of the halo
  virial temperature (solid, coloured lines; see legend for model
  details). The other model parameter, $f_{\rm duty}$, is fixed at
  $10^{-1}$. The profiles of observed CC (dashed grey lines) and NCC
  (solid black lines) clusters in the REXCESS sample (PAP10) are also
  displayed for comparison.}
\label{fig:coolftemp}
\end{figure}

The effect of varying the parameter $f_{\rm temp}$ on the entropy
profile of cluster C1 is shown in Figure \ref{fig:coolftemp}. The
values of $f_{\rm temp}$ we consider are $1$, $2.5$, $4.5$ and $20$,
fixing $f_{\rm duty}=0.1$ in each case. Cluster C1 is a relaxed system
that has not recently undergone any major mergers (half of its mass
was in place at $z\approx 0.8$), so it is a prime candidate for
developing a CC. It is evident that, as $f_{\rm temp}$ is decreased
from $20$ to $1$, the entropy profile steepens, becoming progressively
more like that of a CC cluster. This trend can be explained as
follows. For small values of $f_{\rm temp}$, cold particles in cluster
cores that have received an additional energy input are able to
radiate away this energy more quickly than when $f_{\rm temp}$ is
large, because they have not been heated to such a high temperature
and thus their cooling time is shorter. Therefore, as $f_{\rm temp}$
is decreased, the amount of cool, dense gas in cluster cores
increases, leading to a lower central entropy and a steeper profile.

As mentioned above, an important issue to address is whether our
heating model is energetically plausible. To quantify this, we define
$f_{\rm energy}$ as the ratio of the amount of extra heat energy
supplied to that originally available from the SA model over the
course of the simulation. We want $f_{\rm energy}$ to be as small as
possible at $z=0$. For $f_{\rm temp}=1$, $2.5$, $4.5$ and $20$, we
have $f_{\rm energy}\approx 0.6$, $0.15$, $0.18$ and $0.5$,
respectively, so we discard the models with $f_{\rm heat}=1$ and $20$
on energetic grounds. Small values of $f_{\rm heat}$ ($f_{\rm
  heat}=1$) lead to a large extra energy input because heated gas is
able to cool down relatively quickly in core regions, and is then
heated again, so many extra heating events are required over the
formation history of a cluster. Conversely, when $f_{\rm heat}$ is
large ($f_{\rm heat}=20$), few extra heating events are required
because any cold gas is heated to such high temperatures that its
cooling time becomes very long. However, the large amounts of energy
needed to heat gas to such high temperatures mean that $f_{\rm
  energy}$ is again large.

For the remainder of this section, we choose the model with $(f_{\rm
  rad}, f_{\rm duty}, f_{\rm heat})=(1,10^{-1},2.5)$ as our fiducial
model since this yields a CC-like entropy profile for cluster C1, yet
only requires an extra energy input of $\sim 15\%$ of that available
from the SA model. We have simulated all $25$ clusters in our sample
with this model, and we now examine the predicted thermal properties
of the ICM.

\subsection{Thermal properties of the ICM}

\begin{figure}
\includegraphics[width=85mm]{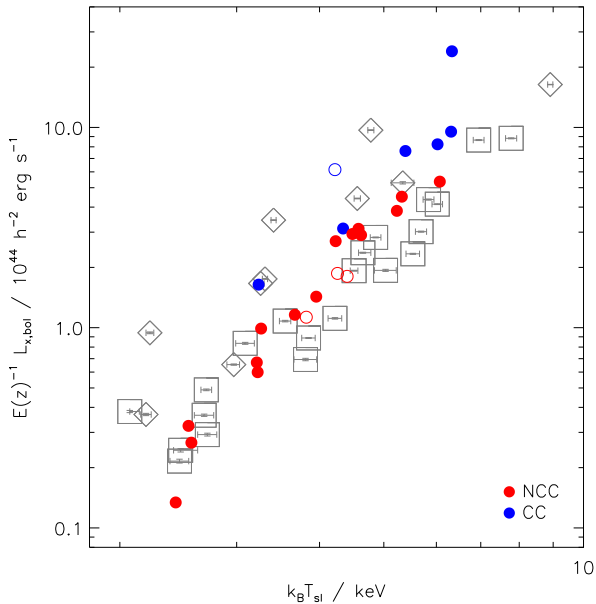}
\caption{The X-ray luminosity-temperature scaling relation predicted
  by our fiducial stochastic AGN feedback model with metal-dependent
  radiative cooling and additional heating of cold gas in cluster
  cores. X-ray properties are computed within $r_{500}$. CC (NCC)
  clusters in our sample are shown as blue (red) circles, while filled
  (open) symbols denote that a cluster is a relaxed (disturbed)
  system. For comparative purposes, we also plot observational data
  for CC (diamonds) and NCC (squares) clusters in the REXCESS sample
  (PCA09).}
\label{fig:coolLT}
\end{figure}

Figure \ref{fig:coolLT} shows our X-ray luminosity-temperature
relation, where both luminosity and spectroscopic-like temperature
have been computed within $r_{500}$. Blue (red) circles represent CC
(NCC) clusters, and filled (open) symbols correspond to relaxed
(disturbed) systems. We classify objects as CC clusters if they are
scattered above the mean observed relation for NCC clusters in the
REXCESS (PAP10) by more than $1\sigma$; $7$ of our $25$ objects
satisfy this criterion. For comparative purposes, we also show
observational data for CC (diamonds) and NCC (squares) clusters in the
REXCESS.

Our predicted relation is not a perfect match to the observational
data, in the sense that the slope appears steeper and there are no
low-temperature systems with a high luminosity, although this could be
a selection effect. However, the salient point is that we are able to
generate both CC and NCC systems with a single feedback model, a feat
that is notoriously difficult with self-consistent hydrodynamical
simulations. This encouraging result warrants further development of
our model in future work.

\begin{figure}
\includegraphics[width=85mm]{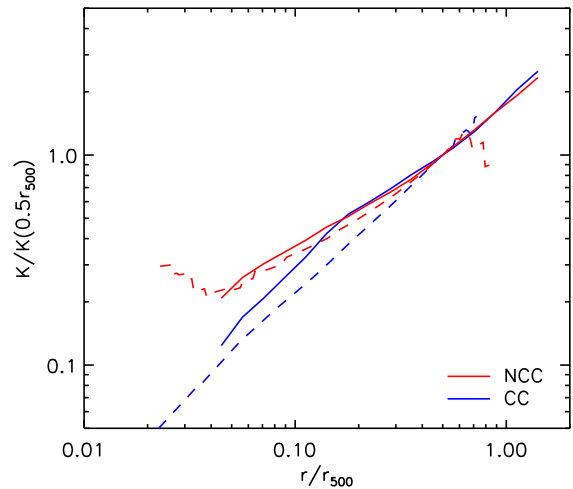}
\caption{\changed{Mean entropy profiles for clusters.  The upper, red lines correspond to
  NCC systems, and the lower, blue lines to CC systems.  The solid lines are
  model clusters resimulated with our fiducial stochastic AGN feedback model,
  plus metal-dependent radiative cooling and additional heating of cold gas in
  cluster cores. The dashed lines are observed profiles of clusters in the
  REXCESS sample (PAP10) that straddle the same mass range.}}
\label{fig:coolKprof}
\end{figure}

\begin{figure}
\includegraphics[width=85mm]{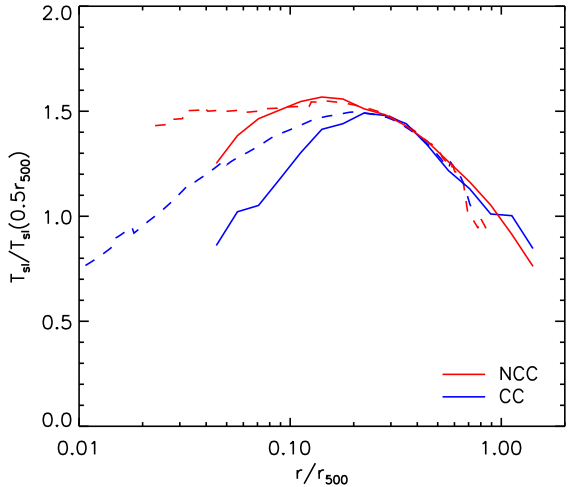}
\caption{\changed{Mean spectroscopic-like temperature profiles for clusters.  The upper,
  red lines correspond to NCC systems, and the lower, blue lines to CC systems.
  The solid lines are model clusters resimulated with our fiducial stochastic
  AGN feedback model, plus metal-dependent radiative cooling and additional
  heating of cold gas in cluster cores. The dashed lines are observed profiles
  of clusters in the REXCESS sample (PAP10) that straddle the same mass range.}}
\label{fig:coolTprof}
\end{figure}

\changed{
The mean entropy and spectroscopic-like temperature profiles of all 25
clusters in our sample are displayed in Figures \ref{fig:coolKprof}
and \ref{fig:coolTprof}, respectively. The profiles of CC (NCC)
clusters are shown by blue (red) lines, with solid (dashed) lines
corresponding to simulated (observed) systems. Note that the dispersion of
individual clusters about these mean relations is quite large, especially for
the temperature profiles, so that it would not be possible to look at a
paricular profile and classify it with certainty as either NCC or CC, according
to our definition above.

There are a number of diferences between the simulated and observed profiles.
Firstly, both the NCC and CC simulated temperature profiles are too low in the
cluster cores.  The entropy profile of simulated CC clusters has the correct
slope, but too high a normalisation below 0.3\,$r_{500}$.  Finally, the
simulated NCC entropy profile shows no sign of flattening at the smallest
radii.  These features all suggest that our heating model is far from perfect
and that perhaps we should target additional heating at not just the coldest
gas.  Nonetheless, it is pleasing that this first attempt should at least lead
to a distinct separation in the mean profiles of the two classes of cluster.
}

\begin{figure}
\includegraphics[width=85mm]{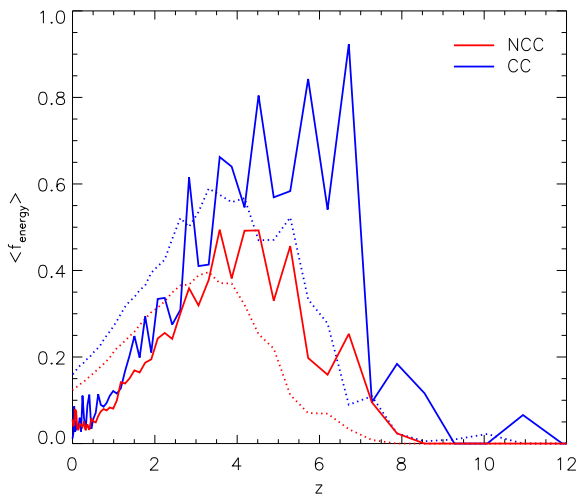}
\caption{Evolution of the ratio of the additional energy input
  required to offset radiative cooling in cluster cores to that
  available from SN and AGN feedback, averaged over all $25$ clusters
  in our sample. We have split our sample into CC (blue) and NCC (red)
  clusters. The solid (dotted) lines show the differential
  (cumulative) evolution.}
\label{fig:extranrg}
\end{figure}

Finally, we assess the energy requirements of our model. In Figure
\ref{fig:extranrg}, we show the ratio $f_{\rm energy}$ as a function
of redshift, averaged over all $25$ clusters. The solid (dotted) lines
show the differential (cumulative) evolution, and again we have
divided our sample into CC (blue lines) and NCC (red lines)
systems. The first point to note is that, at $z=0$, the average total
extra energy input is $\sim 15\%$ of that available from the SA model
(this is actually true for all but one of our objects, which has
$f_{\rm energy}\approx 0.25$), which is reasonable. CC systems require
a larger total energy input, which is to be expected since the gas
cooling time in the central regions of such objects is shorter than in
NCC objects, but the difference is small. Interestingly, for redshifts
$z\lesssim 3$, there is little to distinguish between the two; the
main difference occurs at high redshift, $z\sim 3-8$, where, on
average, much more additional energy is injected into the cores of CC
clusters than NCC clusters. This may be an indication of an earlier
assembly history for CC clusters.

\section{Conclusions}
\label{sec:conc}

In this paper we have used numerical simulations to investigate how
star formation, black hole accretion and the associated feedback from
SNe and AGN heat and enrich intracluster gas. Our primary
objective was to assess how different implementations of these
feedback processes affect the thermal and chemical properties of the
ICM, using a selection of data from X-ray observational studies to
constrain our models.

We have resimulated a sample of $25$ massive galaxy clusters extracted
from the Millennium Simulation. In these simulations, the energy and
metal input into the ICM by SNe and AGN is calculated from a \SAM\ of
galaxy formation, using the hybrid scheme of SHT09. This guarantees
that feedback originates from a realistic galaxy population, whereas
fully self-consistent hydrodynamical simulations often predict
excessive star formation on cluster scales.

Our main achievement has been to develop a new model for AGN feedback
that is both physically motivated and capable of explaining several
fundamental observational properties of clusters. All of the other,
more commonplace, models we have tested fail on one or both of these
points. Our new model is based on stochastic, anisotropic heating of
the ICM, which is motivated by observational evidence that AGN heating
is likely to be directional, rather than isotropic.

Our conclusions are as follows:

\begin{enumerate}
\item Energy input from SN-driven galactic winds has no effect on the entropy
  and metallicity structure of the ICM, regardless of the method used
  to inject energy and metals into the intracluster gas.
\item Simple thermal AGN feedback models all heat the gas excessively
  in the central regions of clusters, generating flat ICM entropy
  profiles that disagree with the observational data. Differences
  between our various models are negligibly small, even though the
  number of particles heated by AGN can vary enormously between models.
\item Kinetic AGN feedback models can reproduce the observed entropy
  profiles of NCC clusters, but only if the wind speed is very high,
  $v_{\rm wind}=20\,000$ km s$^{-1}$, which is possibly unreasonable
  on physical grounds. The success of this model is due to the fact
  that only a small number of particles near the centre of
    the halo are kicked, and the momentum boost they receive is
  sufficient to transport them to cluster outskirts, leaving
  low-entropy gas behind in the core.
\item There are two free parameters in our stochastic heating model:
  $f_{\rm rad}$, which governs the radius of the region (in units of
  the virial radius) about a galaxy in which energy is injected, and
  the fraction, $f_{\rm duty}$, of neighbouring gas particles that are
  heated per AGN duty cycle. The parameter $f_{\rm duty}$ can be
  linked to the opening angle of AGN jets, assuming jets can be simply
  modelled by biconical outflows. Using the observed scaling of X-ray
  luminosity, entropy profile normalisation and shape with temperature
  as constraints, we identified $(f_{\rm rad}, f_{\rm
    duty})=(1,10^{-2})$ as an optimal choice for these parameters. The
  choice $f_{\rm duty}=10^{-2}$ corresponds to a jet opening angle of
  roughly $16^{\circ}$. This is our fiducial model for AGN feedback.
\item Our fiducial stochastic heating model is able to explain both
  the thermal and chemical properties of intracluster gas, at least
  for NCC systems. We obtain a good match to several key pieces of
  observational data: the normalisation and shape of entropy profiles,
  the X-ray luminosity-temperature scaling relation, and the shape of
  metallicity profiles. The model is successful for the same reason as
  the kinetic AGN feedback model with $v_{\rm wind}=20\,000$ km
  s$^{-1}$, but has the advantage of being physically motivated.
\item Reproducing the observed abundance gradient in NCC clusters
  requires that metals ejected from galaxies are distributed
  throughout the entire halo. Injecting metals in a concentrated
  fashion leads to a sharp central peak in the metal distribution that
  is not observed. A metal enrichment model where metals are
  distributed throughout the halo is consistent with the
  \SAM\ underlying our simulations, which predicts that over $95\%$ of
  the metals in diffuse gas are accreted, rather than being produced
  by the central galaxy of the halo.
\item AGN heating causes a flattening of ICM metallicity profiles, but
  the effect is small in our fiducial model. This is because only a
  small fraction of the particles in the core are heated by AGN, and
  they receive a sufficiently large entropy boost to escape the
  central regions of clusters, leaving the majority of metal-rich
  material behind in the core.
\item With the addition of metal-dependent radiative cooling, our
  stochastic AGN feedback model is capable of producing CC and NCC
  systems, avoiding catastrophic over-cooling, but only if we assume
  that additional energy is injected into cold gas in cluster cores to
  offset radiative losses. The justification for this simple model is
  that we expect a mismatch between gas cooling rates in the SA model
  and hydrodynamical simulations. The amount of extra energy typically
  required is $\sim 15\%$ of that available from SN and AGN feedback
  over the formation history of a cluster.
\item \changed{As Figures~\ref{fig:coolKprof} and \ref{fig:coolTprof}
  illustrate, our cooling model is far from a perfect match to the
  observed entropy and temperature profiles in cluster cores.
  Nonetheless, they do show a distinction between CC and NCC clusters.}
\end{enumerate}

To keep the model for the ICM as simple as possible, we have neglected a number
of physical processes in this paper.  While we do not expect any of these to
make a major contribution outside the core of the cluster, it is possible that
their cumulative effect could be important. This should be investigated further
in future extensions of this work:
\begin{itemize}
\item Magnetic fields and cosmic rays will help to provide extra
  pressure support in the ICM.  They can be generated both by mergers
  and by AGN activity.  However, observations suggest that neither
  makes a dominant contribution to clusters, excpet perhaps in the
  core regions.  \citet{Bru11} summarises the current state of play
  for cosmic rays: gamma ray observations from Fermi
  \citep{Fermi10,JeP10} limit the energy density of comsic rays to
  less than a few hundredths of that of the thermal energy of the ICM.
  We note that the qualitative effect on the cluster gas density
  profile of the inclusion of cosmic rays differs between AMR
  \citep{VBG12} and SPH \citep{JSE08} simulations, but the effect on
  the density and temperature profiles of clusters is minor in each
  case.

  Magnetic field strengths in the cores of clusters range from a few
  to a few tens of $\mu$G \citep[e.g.~][and references
    therein]{BFM10,VMG12}.  Measurements for the cluster as a whole
  are hard to make and generally involved a degree of modelling.  The
  observations have been reviewed by \citet{BFM10a} and indicate typical
  values of 1--2\,$\mu$G.  This agrees with theoretical estimates from
  \citet{KSC11}.  At this level, the magnetic energy density will be
  only a minor contributor (of order a percent) to the total energy
  density of the ICM.

\item Conduction has been investigated by many authors, principally as
  a way of overcoming the over-cooling problem in cluster cores
  \citep[e.g.~][]{GuO09,PQS10,RuO10,RuO11,PCQ12}.  Direct evidence for
  conduction is however, by its very nature, almost impossible to
  achieve.  All we have are upper limits based on the existence of
  large temperature gradients surrounding clumps of hot or cold gas
  \citep[e.g.~][]{GXG09,PWS10,DVR12,RMS12}.  It is possible that these
  clumps are surrounded by magnetic sheaths that limit conduction
  across the interface: for the purposes of this paper, what is
  important is the degree by which magnetic fields would suppress
  large-scale conduction between the core and the cluster outskirts,
  and this will depend upon the relative importance of ordering by
  convective motions and the stirring by galaxies and infalling
  substructure.

  Using the observed density and temperature profiles of
  \citet{VKF06}, and assuming conduction at the Spitzer rate, then it
  is possible to estimate the maximum rate at which gas can be heated
  or cooled.  This can be quite large in the cluster core, but is of
  order (10\,Gyr)$^{-1}$ for $r>0.1r_{500}$ (for a 3\,keV cluster).
  Given that some suppression below the Spitzer rate is likely, then
  conductive heat transport at these radii will be minor, but perhaps not
  completely negligible.  We note that the effect would be to heat gas
  within the core and cool gas at larger radii, thus flattening the
  entropy gradient even further and reinforcing the arguments in this
  paper.

\end{itemize}

We have demonstrated that our fiducial stochastic heating model can
explain several important observational properties of massive
clusters, at least for those systems without an X-ray bright CC. With
the inclusion of metal-dependent radiative cooling and a simple
prescription for additional heating of gas in the central regions of
clusters, we have taken our first steps to being able to produce both
CC and NCC systems with a single model. We are currently undertaking a
hydrodynamical simulation of the full Millennium volume ($500 h^{-3}$
Mpc$^3$) with these models. The aim is to produce a large,
publicly-available sample of galaxy groups and clusters whose
properties are consistent with the available X-ray data. An example of
an important application of such a sample would be modelling the
selection functions of X-ray surveys (e.g. \citealt{SVL09}). This is
essential to exploit the full power of clusters as cosmological probes
of the expansion history of the Universe.

\section*{Acknowledgements}

We are grateful to G.~W.~Pratt and K.~Matsushita for supplying us with
their observational data. We also thank V.~Springel for providing the
merger tree software, the \lgalaxies\ Steering Committee for making
the model available to us, and A.~Jenkins for the code to
generate initial conditions. The calculations for this paper were
performed on the ICC Cosmology Machine, which is part of the DiRAC
Facility jointly funded by the Science and Technology Facilities
Council, the Large Facilities Capital Fund of BIS, and Durham
University. Finally, we acknowledge support from the Science and
Technology Facilities Council (grant number ST/I000976/1).

\bibliographystyle{mn2e} \bibliography{bibliography}

\begin{thebibliography}{135}
\expandafter\ifx\csname natexlab\endcsname\relax\def\natexlab#1{#1}\fi

\bibitem[{{Ackermann} \& {et al.}(2010)}]{Fermi10}
{Ackermann} M., {et al.}, 2010, ApJ, 717, 71

\bibitem[{{Akritas} \& {Bershady}(1996)}]{AKB96}
{Akritas} M.~G., {Bershady} M.~A., 1996, ApJ, 470, 706

\bibitem[{{Alexander} {et~al.}(2010){Alexander}, {Swinbank}, {Smail},
  {McDermid}, \& {Nesvadba}}]{ASS10}
{Alexander} D.~M., {Swinbank} A.~M., {Smail} I., {McDermid} R., {Nesvadba}
  N.~P.~H., 2010, MNRAS, 402, 2211

\bibitem[{Arnaud {et~al.}(2010)Arnaud, Pratt, Piffaretti, Boehringer, Croston,
  \& Pointecouteau}]{APP09}
Arnaud M., Pratt G., Piffaretti R., Boehringer H., Croston J., Pointecouteau
  E., 2010, A\&A, 517, 92

\bibitem[{{Baldi} {et~al.}(2007){Baldi}, {Ettori}, {Mazzotta}, {Tozzi}, \&
  {Borgani}}]{BEM07}
{Baldi} A., {Ettori} S., {Mazzotta} P., {Tozzi} P., {Borgani} S., 2007, ApJ,
  666, 835

\bibitem[{{Balogh} {et~al.}(2008){Balogh}, {McCarthy}, {Bower}, \&
  {Eke}}]{BMBE08}
{Balogh} M.~L., {McCarthy} I.~G., {Bower} R.~G., {Eke} V.~R., 2008, MNRAS, 385,
  1003

\bibitem[{{Balogh} {et~al.}(2001){Balogh}, {Pearce}, {Bower}, \& {Kay}}]{BPB01}
{Balogh} M.~L., {Pearce} F.~R., {Bower} R.~G., {Kay} S.~T., 2001, MNRAS, 326,
  1228

\bibitem[{{Bhattacharya} {et~al.}(2008){Bhattacharya}, {Di Matteo}, \&
  {Kosowsky}}]{BDK08}
{Bhattacharya} S., {Di Matteo} T., {Kosowsky} A., 2008, MNRAS, 389, 34

\bibitem[{{B{\^i}rzan} {et~al.}(2004){B{\^i}rzan}, {Rafferty}, {McNamara},
  {Wise}, \& {Nulsen}}]{BRM04}
{B{\^i}rzan} L., {Rafferty} D.~A., {McNamara} B.~R., {Wise} M.~W., {Nulsen}
  P.~E.~J., 2004, ApJ, 607, 800

\bibitem[{{Blanton} {et~al.}(2001){Blanton}, {Sarazin}, {McNamara}, \&
  {Wise}}]{BSM01}
{Blanton} E.~L., {Sarazin} C.~L., {McNamara} B.~R., {Wise} M.~W., 2001, ApJ,
  558, L15

\bibitem[{{B{\"o}hringer} \& {Werner}(2010)}]{BOW10}
{B{\"o}hringer} H., {Werner} N., 2010, A\&AR, 18, 127

\bibitem[{{Bonafede} {et~al.}(2010){Bonafede}, {Feretti}, {Murgia}, {Govoni},
  {Giovannini}, {Dallacasa}, {Dolag}, \& {Taylor}}]{BFM10}
{Bonafede} A., {Feretti} L., {Murgia} M., {Govoni} F., {Giovannini} G.,
  {Dallacasa} D., {Dolag} K., {Taylor} G.~B., 2010, A\&A, 513, 30

\bibitem[{Bonafede {et~al.}(2010)Bonafede, Feretti, Murgia, Govoni, Giovannini,
  \& Vacca}]{BFM10a}
Bonafede A., Feretti L., Murgia M., Govoni F., Giovannini G., Vacca V., 2010,
  ArXiv e-prints: 1009.1233

\bibitem[{{Booth} \& {Schaye}(2009)}]{BOS09}
{Booth} C.~M., {Schaye} J., 2009, MNRAS, 398, 53

\bibitem[{{Borgani} \& {Kravtsov}(2009)}]{BOK09}
{Borgani} S., {Kravtsov} A., 2009, ArXiv e-prints: 0906.4370

\bibitem[{{Borgani} {et~al.}(2004){Borgani}, {Murante}, {Springel}, {Diaferio},
  {Dolag}, {Moscardini}, {Tormen}, {Tornatore}, \& {Tozzi}}]{BMS04}
{Borgani} S., {Murante} G., {Springel} V., {Diaferio} A., {Dolag} K.,
  {Moscardini} L., {Tormen} G., {Tornatore} L., {Tozzi} P., 2004, MNRAS, 348,
  1078

\bibitem[{{Brook} {et~al.}(2004){Brook}, {Kawata}, {Gibson}, \&
  {Flynn}}]{BKG04}
{Brook} C.~B., {Kawata} D., {Gibson} B.~K., {Flynn} C., 2004, MNRAS, 349, 52

\bibitem[{{Br{\"u}ggen}(2003)}]{BRU03}
{Br{\"u}ggen} M., 2003, ApJ, 592, 839

\bibitem[{{Br{\"u}ggen} {et~al.}(2002){Br{\"u}ggen}, {Kaiser}, {Churazov}, \&
  {En{\ss}lin}}]{BKC02}
{Br{\"u}ggen} M., {Kaiser} C.~R., {Churazov} E., {En{\ss}lin} T.~A., 2002,
  MNRAS, 331, 545

\bibitem[{{Br{\"u}ggen} \& {Scannapieco}(2009)}]{BRS09}
{Br{\"u}ggen} M., {Scannapieco} E., 2009, MNRAS, 398, 548

\bibitem[{Brunetti(2011)}]{Bru11}
Brunetti G., 2011, Memorie della Societa Astronomica Italiana, 82, 515

\bibitem[{{Cavagnolo} {et~al.}(2009){Cavagnolo}, {Donahue}, {Voit}, \&
  {Sun}}]{CDV09}
{Cavagnolo} K.~W., {Donahue} M., {Voit} G.~M., {Sun} M., 2009, ApJ Supp., 182,
  12

\bibitem[{{Chartas} {et~al.}(2003){Chartas}, {Brandt}, \& {Gallagher}}]{CBG03}
{Chartas} G., {Brandt} W.~N., {Gallagher} S.~C., 2003, ApJ, 595, 85

\bibitem[{{Churazov} {et~al.}(2001){Churazov}, {Br{\"u}ggen}, {Kaiser},
  {B{\"o}hringer}, \& {Forman}}]{CBK01}
{Churazov} E., {Br{\"u}ggen} M., {Kaiser} C.~R., {B{\"o}hringer} H., {Forman}
  W., 2001, ApJ, 554, 261

\bibitem[{{Cora}(2006)}]{COR06}
{Cora} S.~A., 2006, MNRAS, 368, 1540

\bibitem[{{Cora} {et~al.}(2008){Cora}, {Tornatore}, {Tozzi}, \&
  {Dolag}}]{CTT08}
{Cora} S.~A., {Tornatore} L., {Tozzi} P., {Dolag} K., 2008, MNRAS, 386, 96

\bibitem[{{Crenshaw} {et~al.}(2003){Crenshaw}, {Kraemer}, \& {George}}]{CKG03}
{Crenshaw} D.~M., {Kraemer} S.~B., {George} I.~M., 2003, ARA\&A, 41, 117

\bibitem[{{Croton} {et~al.}(2006){Croton}, {Springel}, {White}, {De Lucia},
  {Frenk}, {Gao}, {Jenkins}, {Kauffmann}, {Navarro}, \& {Yoshida}}]{CSW06}
{Croton} D.~J., {Springel} V., {White} S.~D.~M., {De Lucia} G., {Frenk} C.~S.,
  {Gao} L., {Jenkins} A., {Kauffmann} G., {Navarro} J.~F., {Yoshida} N., 2006,
  MNRAS, 365, 11

\bibitem[{{Dalla Vecchia} {et~al.}(2004){Dalla Vecchia}, {Bower}, {Theuns},
  {Balogh}, {Mazzotta}, \& {Frenk}}]{DBT04}
{Dalla Vecchia} C., {Bower} R.~G., {Theuns} T., {Balogh} M.~L., {Mazzotta} P.,
  {Frenk} C.~S., 2004, MNRAS, 355, 995

\bibitem[{{Dalla Vecchia} \& {Schaye}(2008)}]{DVS08}
{Dalla Vecchia} C., {Schaye} J., 2008, MNRAS, 387, 1431

\bibitem[{{Dav{\'e}} {et~al.}(2008){Dav{\'e}}, {Oppenheimer}, \&
  {Sivanandam}}]{DOS08}
{Dav{\'e}} R., {Oppenheimer} B.~D., {Sivanandam} S., 2008, MNRAS, 391, 110

\bibitem[{{Davis} {et~al.}(1985){Davis}, {Efstathiou}, {Frenk}, \&
  {White}}]{DEF85}
{Davis} M., {Efstathiou} G., {Frenk} C.~S., {White} S.~D.~M., 1985, ApJ, 292,
  371

\bibitem[{{De Grandi} \& {Molendi}(2001)}]{DEM01}
{De Grandi} S., {Molendi} S., 2001, ApJ, 551, 153

\bibitem[{{De Lucia} \& {Blaizot}(2007)}]{DLB07}
{De Lucia} G., {Blaizot} J., 2007, MNRAS, 375, 2

\bibitem[{{de Plaa} {et~al.}(2010){de Plaa}, {Werner}, {Simionescu}, {Kaastra},
  {Grange}, \& {Vink}}]{PWS10}
{de Plaa} J., {Werner} N., {Simionescu} A., {Kaastra} J.~S., {Grange} Y.~G.,
  {Vink} J., 2010, A\&A, 523, A81

\bibitem[{{Di Matteo} {et~al.}(2008){Di Matteo}, {Colberg}, {Springel},
  {Hernquist}, \& {Sijacki}}]{DCS08}
{Di Matteo} T., {Colberg} J., {Springel} V., {Hernquist} L., {Sijacki} D.,
  2008, ApJ, 676, 33

\bibitem[{{Di Matteo} {et~al.}(2005){Di Matteo}, {Springel}, \&
  {Hernquist}}]{DSH05}
{Di Matteo} T., {Springel} V., {Hernquist} L., 2005, Nat., 433, 604

\bibitem[{{Dong} {et~al.}(2010){Dong}, {Rasmussen}, \& {Mulchaey}}]{DRM10}
{Dong} R., {Rasmussen} J., {Mulchaey} J.~S., 2010, ApJ, 712, 883

\bibitem[{{Dubois} {et~al.}(2011){Dubois}, {Devriendt}, {Teyssier}, \&
  {Slyz}}]{DDT11}
{Dubois} Y., {Devriendt} J., {Teyssier} R., {Slyz} A., 2011, MNRAS, 1345

\bibitem[{{Dubois} \& {Teyssier}(2008)}]{DUT08}
{Dubois} Y., {Teyssier} R., 2008, A\&A, 477, 79

\bibitem[{{Dunn} {et~al.}(2010){Dunn}, {Bautista}, {Arav}, {Moe}, {Korista},
  {Costantini}, {Benn}, {Ellison}, \& {Edmonds}}]{DBA10}
{Dunn} J.~P., {Bautista} M., {Arav} N., {Moe} M., {Korista} K., {Costantini}
  E., {Benn} C., {Ellison} S., {Edmonds} D., 2010, ApJ, 709, 611

\bibitem[{{Fabian} {et~al.}(2006){Fabian}, {Sanders}, {Taylor}, {Allen},
  {Crawford}, {Johnstone}, \& {Iwasawa}}]{FST06}
{Fabian} A.~C., {Sanders} J.~S., {Taylor} G.~B., {Allen} S.~W., {Crawford}
  C.~S., {Johnstone} R.~M., {Iwasawa} K., 2006, MNRAS, 366, 417

\bibitem[{{Fabjan} {et~al.}(2010){Fabjan}, {Borgani}, {Tornatore}, {Saro},
  {Murante}, \& {Dolag}}]{FBT10}
{Fabjan} D., {Borgani} S., {Tornatore} L., {Saro} A., {Murante} G., {Dolag} K.,
  2010, MNRAS, 401, 1670

\bibitem[{{Finoguenov} {et~al.}(2000){Finoguenov}, {David}, \&
  {Ponman}}]{FDP00}
{Finoguenov} A., {David} L.~P., {Ponman} T.~J., 2000, ApJ, 544, 188

\bibitem[{{Forman} {et~al.}(2005){Forman}, {Nulsen}, {Heinz}, {Owen}, {Eilek},
  {Vikhlinin}, {Markevitch}, {Kraft}, {Churazov}, \& {Jones}}]{FNH05}
{Forman} W., {Nulsen} P., {Heinz} S., {Owen} F., {Eilek} J., {Vikhlinin} A.,
  {Markevitch} M., {Kraft} R., {Churazov} E., {Jones} C., 2005, ApJ, 635, 894

\bibitem[{{Ganguly} \& {Brotherton}(2008)}]{GAB08}
{Ganguly} R., {Brotherton} M.~S., 2008, ApJ, 672, 102

\bibitem[{Gaspari {et~al.}(2012)Gaspari, Ruszkowski, \& Sharma}]{GRS12}
Gaspari M., Ruszkowski M., Sharma P., 2012, ApJ, 746, 94

\bibitem[{{Gastaldello} {et~al.}(2009){Gastaldello}, {Buote}, {Temi},
  {Brighenti}, {Mathews}, \& {Ettori}}]{GBT09}
{Gastaldello} F., {Buote} D.~A., {Temi} P., {Brighenti} F., {Mathews} W.~G.,
  {Ettori} S., 2009, ApJ, 693, 43

\bibitem[{{Giacintucci} {et~al.}(2011){Giacintucci}, {O'Sullivan}, {Vrtilek},
  {David}, {Raychaudhury}, {Venturi}, {Athreya}, {Clarke}, {Murgia},
  {Mazzotta}, {Gitti}, {Ponman}, {Ishwara-Chandra}, {Jones}, \&
  {Forman}}]{GOV11}
{Giacintucci} S., {O'Sullivan} E., {Vrtilek} J., {David} L.~P., {Raychaudhury}
  S., {Venturi} T., {Athreya} R.~M., {Clarke} T.~E., {Murgia} M., {Mazzotta}
  P., {Gitti} M., {Ponman} T., {Ishwara-Chandra} C.~H., {Jones} C., {Forman}
  W.~R., 2011, ApJ, 732, 95

\bibitem[{{Gu} {et~al.}(2009){Gu}, {Xu}, {Gu}, {Wang}, {Zhang}, {Wang}, {Qin},
  {Cui}, \& {Wu}}]{GXG09}
{Gu} L., {Xu} H., {Gu} J., {Wang} Y., {Zhang} Z., {Wang} J., {Qin} Z., {Cui}
  H., {Wu} X.-P., 2009, ApJ, 700, 1161

\bibitem[{{Guo} \& {Oh}(2009)}]{GuO09}
{Guo} F., {Oh} S.~P., 2009, MNRAS, 400, 1992

\bibitem[{{Henriques} \& {Thomas}(2010)}]{HET10}
{Henriques} B.~M.~B., {Thomas} P.~A., 2010, MNRAS, 403, 768

\bibitem[{Jeltema \& Profumo(2010)}]{JeP10}
Jeltema T.~E., Profumo S., 2010, ApJ, 728, 53

\bibitem[{{Jetha} {et~al.}(2008){Jetha}, {Hardcastle}, {Ponman}, \&
  {Sakelliou}}]{JHP08}
{Jetha} N.~N., {Hardcastle} M.~J., {Ponman} T.~J., {Sakelliou} I., 2008, MNRAS,
  391, 1052

\bibitem[{Jubelgas {et~al.}(2008)Jubelgas, Springel, Ensslin, \&
  Pfrommer}]{JSE08}
Jubelgas M., Springel V., Ensslin T.~A., Pfrommer C., 2008, A\&A, 481, 33

\bibitem[{{Katz}(1992)}]{KAT92}
{Katz} N., 1992, ApJ, 391, 502

\bibitem[{{Kawata}(2001)}]{KAW01}
{Kawata} D., 2001, ApJ, 558, 598

\bibitem[{{Kay} {et~al.}(2007){Kay}, {da Silva}, {Aghanim}, {Blanchard},
  {Liddle}, {Puget}, {Sadat}, \& {Thomas}}]{KDA07}
{Kay} S.~T., {da Silva} A.~C., {Aghanim} N., {Blanchard} A., {Liddle} A.~R.,
  {Puget} J.-L., {Sadat} R., {Thomas} P.~A., 2007, MNRAS, 377, 317

\bibitem[{{Kay} {et~al.}(2002){Kay}, {Pearce}, {Frenk}, \& {Jenkins}}]{KPF02}
{Kay} S.~T., {Pearce} F.~R., {Frenk} C.~S., {Jenkins} A., 2002, MNRAS, 330, 113

\bibitem[{{Kirkpatrick} {et~al.}(2011){Kirkpatrick}, {McNamara}, \&
  {Cavagnolo}}]{KMC11}
{Kirkpatrick} C.~C., {McNamara} B.~R., {Cavagnolo} K.~W., 2011, ApJ, 731, L23

\bibitem[{Kunz {et~al.}(2011)Kunz, Schekochihin, Cowley, Binney, \&
  Sanders}]{KSC11}
Kunz M.~W., Schekochihin A.~A., Cowley S.~C., Binney J.~J., Sanders J.~S.,
  2011, MNRAS, 410, 2446

\bibitem[{{Leccardi} \& {Molendi}(2008)}]{LEM08}
{Leccardi} A., {Molendi} S., 2008, A\&A, 487, 461

\bibitem[{{Leccardi} {et~al.}(2010){Leccardi}, {Rossetti}, \&
  {Molendi}}]{LRM10}
{Leccardi} A., {Rossetti} M., {Molendi} S., 2010, A\&A, 510, 82

\bibitem[{{Lin} {et~al.}(2003){Lin}, {Mohr}, \& {Stanford}}]{LMS03}
{Lin} Y.-T., {Mohr} J.~J., {Stanford} S.~A., 2003, ApJ, 591, 749

\bibitem[{{Lloyd-Davies} {et~al.}(2000){Lloyd-Davies}, {Ponman}, \&
  {Cannon}}]{LPC00}
{Lloyd-Davies} E.~J., {Ponman} T.~J., {Cannon} D.~B., 2000, MNRAS, 315, 689

\bibitem[{{Martin}(2005)}]{MAR05}
{Martin} C.~L., 2005, ApJ, 621, 227

\bibitem[{{Matsushita}(2011)}]{MAT11}
{Matsushita} K., 2011, A\&A, 527, 134

\bibitem[{{Maughan} {et~al.}(2008){Maughan}, {Jones}, {Forman}, \& {Van
  Speybroeck}}]{MJF08}
{Maughan} B.~J., {Jones} C., {Forman} W., {Van Speybroeck} L., 2008, ApJ Supp.,
  174, 117, {(MJF08)}

\bibitem[{{McCarthy} {et~al.}(2010){McCarthy}, {Schaye}, {Ponman}, {Bower},
  {Booth}, {Vecchia}, {Crain}, {Springel}, {Theuns}, \& {Wiersma}}]{MSP10}
{McCarthy} I.~G., {Schaye} J., {Ponman} T.~J., {Bower} R.~G., {Booth} C.~M.,
  {Vecchia} C.~D., {Crain} R.~A., {Springel} V., {Theuns} T., {Wiersma}
  R.~P.~C., 2010, MNRAS, 740

\bibitem[{{McDonald} {et~al.}(2012){McDonald}, {Veilleux}, \& {Rupke}}]{DVR12}
{McDonald} M., {Veilleux} S., {Rupke} D.~S.~N., 2012, ApJ, 746, 153

\bibitem[{{McNamara} \& {Nulsen}(2007)}]{MCN07}
{McNamara} B.~R., {Nulsen} P.~E.~J., 2007, ARA\&A, 45, 117

\bibitem[{{McNamara} {et~al.}(2005){McNamara}, {Nulsen}, {Wise}, {Rafferty},
  {Carilli}, {Sarazin}, \& {Blanton}}]{MNW05}
{McNamara} B.~R., {Nulsen} P.~E.~J., {Wise} M.~W., {Rafferty} D.~A., {Carilli}
  C., {Sarazin} C.~L., {Blanton} E.~L., 2005, Nat., 433, 45

\bibitem[{{Mihos} \& {Hernquist}(1994)}]{MIH94}
{Mihos} J.~C., {Hernquist} L., 1994, ApJ, 437, 611

\bibitem[{{Million} {et~al.}(2010){Million}, {Werner}, {Simionescu}, {Allen},
  {Nulsen}, {Fabian}, {B{\"o}hringer}, \& {Sanders}}]{MWS10}
{Million} E.~T., {Werner} N., {Simionescu} A., {Allen} S.~W., {Nulsen}
  P.~E.~J., {Fabian} A.~C., {B{\"o}hringer} H., {Sanders} J.~S., 2010, MNRAS,
  407, 2046

\bibitem[{{Moll} {et~al.}(2007){Moll}, {Schindler}, {Domainko}, {Kapferer},
  {Mair}, {van Kampen}, {Kronberger}, {Kimeswenger}, \& {Ruffert}}]{MSD07}
{Moll} R., {Schindler} S., {Domainko} W., {Kapferer} W., {Mair} M., {van
  Kampen} E., {Kronberger} T., {Kimeswenger} S., {Ruffert} M., 2007, A\&A, 463,
  513

\bibitem[{{Morandi} \& {Ettori}(2007)}]{MOE07}
{Morandi} A., {Ettori} S., 2007, MNRAS, 380, 1521

\bibitem[{{Mori} {et~al.}(1997){Mori}, {Yoshii}, {Tsujimoto}, \&
  {Nomoto}}]{MYT97}
{Mori} M., {Yoshii} Y., {Tsujimoto} T., {Nomoto} K., 1997, ApJ, 478, L21

\bibitem[{{Morita} {et~al.}(2006){Morita}, {Ishisaki}, {Yamasaki}, {Ota},
  {Kawano}, {Fukazawa}, \& {Ohashi}}]{MIY06}
{Morita} U., {Ishisaki} Y., {Yamasaki} N.~Y., {Ota} N., {Kawano} N., {Fukazawa}
  Y., {Ohashi} T., 2006, PASJ, 58, 719

\bibitem[{{Muanwong} {et~al.}(2002){Muanwong}, {Thomas}, {Kay}, \&
  {Pearce}}]{MTK02}
{Muanwong} O., {Thomas} P.~A., {Kay} S.~T., {Pearce} F.~R., 2002, MNRAS, 336,
  527

\bibitem[{{Nagai} {et~al.}(2007){Nagai}, {Kravtsov}, \& {Vikhlinin}}]{NKV07}
{Nagai} D., {Kravtsov} A.~V., {Vikhlinin} A., 2007, ApJ, 668, 1

\bibitem[{{Navarro} \& {White}(1993)}]{NAW93}
{Navarro} J.~F., {White} S.~D.~M., 1993, MNRAS, 265, 271

\bibitem[{{Nesvadba} {et~al.}(2008){Nesvadba}, {Lehnert}, {De Breuck},
  {Gilbert}, \& {van Breugel}}]{NLD08}
{Nesvadba} N.~P.~H., {Lehnert} M.~D., {De Breuck} C., {Gilbert} A.~M., {van
  Breugel} W., 2008, A\&A, 491, 407

\bibitem[{{Nesvadba} {et~al.}(2006){Nesvadba}, {Lehnert}, {Eisenhauer},
  {Gilbert}, {Tecza}, \& {Abuter}}]{NLE06}
{Nesvadba} N.~P.~H., {Lehnert} M.~D., {Eisenhauer} F., {Gilbert} A., {Tecza}
  M., {Abuter} R., 2006, ApJ, 650, 693

\bibitem[{{Okamoto} {et~al.}(2005){Okamoto}, {Eke}, {Frenk}, \&
  {Jenkins}}]{OEF05}
{Okamoto} T., {Eke} V.~R., {Frenk} C.~S., {Jenkins} A., 2005, MNRAS, 363, 1299

\bibitem[{{Oppenheimer} \& {Dav{\'e}}(2006)}]{OPD06}
{Oppenheimer} B.~D., {Dav{\'e}} R., 2006, MNRAS, 373, 1265

\bibitem[{Parrish {et~al.}(2012)Parrish, McCourt, Quataert, \& Sharma}]{PCQ12}
Parrish I.~J., McCourt M., Quataert E., Sharma P., 2012, MNRAS, 422, 704

\bibitem[{{Parrish} {et~al.}(2010){Parrish}, {Quataert}, \& {Sharma}}]{PQS10}
{Parrish} I.~J., {Quataert} E., {Sharma} P., 2010, ApJ, 712, L194

\bibitem[{{Ponman} {et~al.}(1999){Ponman}, {Cannon}, \& {Navarro}}]{PCN99}
{Ponman} T.~J., {Cannon} D.~B., {Navarro} J.~F., 1999, Nat., 397, 135

\bibitem[{{Ponman} {et~al.}(2003){Ponman}, {Sanderson}, \&
  {Finoguenov}}]{PSF03}
{Ponman} T.~J., {Sanderson} A.~J.~R., {Finoguenov} A., 2003, MNRAS, 343, 331

\bibitem[{{Pounds} {et~al.}(2003){Pounds}, {Reeves}, {King}, {Page}, {O'Brien},
  \& {Turner}}]{PRK03}
{Pounds} K.~A., {Reeves} J.~N., {King} A.~R., {Page} K.~L., {O'Brien} P.~T.,
  {Turner} M.~J.~L., 2003, MNRAS, 345, 705

\bibitem[{{Pratt} {et~al.}(2010){Pratt}, {Arnaud}, {Piffaretti},
  {B{\"o}hringer}, {Ponman}, {Croston}, {Voit}, {Borgani}, \& {Bower}}]{PAP10}
{Pratt} G.~W., {Arnaud} M., {Piffaretti} R., {B{\"o}hringer} H., {Ponman}
  T.~J., {Croston} J.~H., {Voit} G.~M., {Borgani} S., {Bower} R.~G., 2010,
  A\&A, 511, A85, {(PAP10)}

\bibitem[{{Pratt} {et~al.}(2006){Pratt}, {Arnaud}, \& {Pointecouteau}}]{PAP06}
{Pratt} G.~W., {Arnaud} M., {Pointecouteau} E., 2006, A\&A, 446, 429

\bibitem[{{Pratt} {et~al.}(2007){Pratt}, {B{\"o}hringer}, {Croston}, {Arnaud},
  {Borgani}, {Finoguenov}, \& {Temple}}]{PBC07}
{Pratt} G.~W., {B{\"o}hringer} H., {Croston} J.~H., {Arnaud} M., {Borgani} S.,
  {Finoguenov} A., {Temple} R.~F., 2007, A\&A, 461, 71

\bibitem[{{Puchwein} {et~al.}(2008){Puchwein}, {Sijacki}, \&
  {Springel}}]{PSS08}
{Puchwein} E., {Sijacki} D., {Springel} V., 2008, ApJ, 687, L53

\bibitem[{{Quilis} {et~al.}(2001){Quilis}, {Bower}, \& {Balogh}}]{QBB01}
{Quilis} V., {Bower} R.~G., {Balogh} M.~L., 2001, MNRAS, 328, 1091

\bibitem[{{Rasmussen} \& {Ponman}(2009)}]{RAP09}
{Rasmussen} J., {Ponman} T.~J., 2009, MNRAS, 399, 239

\bibitem[{{Roediger} {et~al.}(2007){Roediger}, {Br{\"u}ggen}, {Rebusco},
  {B{\"o}hringer}, \& {Churazov}}]{RBR07}
{Roediger} E., {Br{\"u}ggen} M., {Rebusco} P., {B{\"o}hringer} H., {Churazov}
  E., 2007, MNRAS, 375, 15

\bibitem[{{Romeo} {et~al.}(2006){Romeo}, {Sommer-Larsen}, {Portinari}, \&
  {Antonuccio-Delogu}}]{RSP06}
{Romeo} A.~D., {Sommer-Larsen} J., {Portinari} L., {Antonuccio-Delogu} V.,
  2006, MNRAS, 371, 548

\bibitem[{{Rossetti} \& {Molendi}(2010)}]{ROM10}
{Rossetti} M., {Molendi} S., 2010, A\&A, 510, 83

\bibitem[{Russell {et~al.}(2012)Russell, McNamara, Sanders, Fabian, Nulsen,
  Canning, Baum, Donahue, Edge, King, \& O'Dea}]{RMS12}
Russell H.~R., McNamara B.~R., Sanders J.~S., Fabian A.~C., Nulsen P. E.~J.,
  Canning R. E.~A., Baum S.~A., Donahue M., Edge A.~C., King L.~J., O'Dea
  C.~P., 2012, ArXiv e-prints: 1202.5320

\bibitem[{{Ruszkowski} \& {Begelman}(2002)}]{RUB02}
{Ruszkowski} M., {Begelman} M.~C., 2002, ApJ, 581, 223

\bibitem[{Ruszkowski \& Oh(2010)}]{RuO11}
Ruszkowski M., Oh S.~P., 2010, MNRAS, 414, 1493

\bibitem[{{Ruszkowski} \& {Oh}(2010)}]{RuO10}
{Ruszkowski} M., {Oh} S.~P., 2010, ApJ, 713, 1332

\bibitem[{{Sahl{\'e}n} {et~al.}(2009){Sahl{\'e}n}, {Viana}, {Liddle}, {Romer},
  {Davidson}, {Hosmer}, {Lloyd-Davies}, {Sabirli}, {Collins}, {Freeman},
  {Hilton}, {Hoyle}, {Kay}, {Mann}, {Mehrtens}, {Miller}, {Nichol}, {Stanford},
  \& {West}}]{SVL09}
{Sahl{\'e}n} M., {Viana} P.~T.~P., {Liddle} A.~R., {Romer} A.~K., {Davidson}
  M., {Hosmer} M., {Lloyd-Davies} E., {Sabirli} K., {Collins} C.~A., {Freeman}
  P.~E., {Hilton} M., {Hoyle} B., {Kay} S.~T., {Mann} R.~G., {Mehrtens} N.,
  {Miller} C.~J., {Nichol} R.~C., {Stanford} S.~A., {West} M.~J., 2009, MNRAS,
  397, 577

\bibitem[{{Sanderson} {et~al.}(2009){Sanderson}, {O'Sullivan}, \&
  {Ponman}}]{SOP09}
{Sanderson} A.~J.~R., {O'Sullivan} E., {Ponman} T.~J., 2009, MNRAS, 395, 764

\bibitem[{{Sanderson} {et~al.}(2006){Sanderson}, {Ponman}, \&
  {O'Sullivan}}]{SPO06}
{Sanderson} A.~J.~R., {Ponman} T.~J., {O'Sullivan} E., 2006, MNRAS, 372, 1496

\bibitem[{{Short} \& {Thomas}(2009)}]{SHT09}
{Short} C.~J., {Thomas} P.~A., 2009, ApJ, 704, 915

\bibitem[{{Short} {et~al.}(2010){Short}, {Thomas}, {Young}, {Pearce},
  {Jenkins}, \& {Muanwong}}]{STY10}
{Short} C.~J., {Thomas} P.~A., {Young} O.~E., {Pearce} F.~R., {Jenkins} A.,
  {Muanwong} O., 2010, MNRAS, 408, 2213

\bibitem[{{Sijacki} \& {Springel}(2006)}]{SIS06}
{Sijacki} D., {Springel} V., 2006, MNRAS, 366, 397

\bibitem[{{Sijacki} {et~al.}(2007){Sijacki}, {Springel}, {di Matteo}, \&
  {Hernquist}}]{SSD07}
{Sijacki} D., {Springel} V., {di Matteo} T., {Hernquist} L., 2007, MNRAS, 380,
  877

\bibitem[{{Springel}(2005)}]{SPR05}
{Springel} V., 2005, MNRAS, 364, 1105

\bibitem[{{Springel} {et~al.}(2005{\natexlab{a}}){Springel}, {Di Matteo}, \&
  {Hernquist}}]{SDH05a}
{Springel} V., {Di Matteo} T., {Hernquist} L., 2005{\natexlab{a}}, MNRAS, 361,
  776

\bibitem[{{Springel} \& {Hernquist}(2003)}]{SPH03}
{Springel} V., {Hernquist} L., 2003, MNRAS, 339, 289

\bibitem[{{Springel} {et~al.}(2005{\natexlab{b}}){Springel}, {White},
  {Jenkins}, {Frenk}, {Yoshida}, {Gao}, {Navarro}, {Thacker}, {Croton},
  {Helly}, {Peacock}, {Cole}, {Thomas}, {Couchman}, {Evrard}, {Colberg}, \&
  {Pearce}}]{SWJ05}
{Springel} V., {White} S.~D.~M., {Jenkins} A., {Frenk} C.~S., {Yoshida} N.,
  {Gao} L., {Navarro} J., {Thacker} R., {Croton} D., {Helly} J., {Peacock}
  J.~A., {Cole} S., {Thomas} P., {Couchman} H., {Evrard} A., {Colberg} J.,
  {Pearce} F., 2005{\natexlab{b}}, Nat., 435, 629

\bibitem[{{Springel} {et~al.}(2001){Springel}, {White}, {Tormen}, \&
  {Kauffmann}}]{SWT01}
{Springel} V., {White} S.~D.~M., {Tormen} G., {Kauffmann} G., 2001, MNRAS, 328,
  726

\bibitem[{{Steidel} {et~al.}(2010){Steidel}, {Erb}, {Shapley}, {Pettini},
  {Reddy}, {Bogosavljevi{\'c}}, {Rudie}, \& {Rakic}}]{SES10}
{Steidel} C.~C., {Erb} D.~K., {Shapley} A.~E., {Pettini} M., {Reddy} N.,
  {Bogosavljevi{\'c}} M., {Rudie} G.~C., {Rakic} O., 2010, ApJ, 717, 289

\bibitem[{{Stinson} {et~al.}(2006){Stinson}, {Seth}, {Katz}, {Wadsley},
  {Governato}, \& {Quinn}}]{SSK06}
{Stinson} G., {Seth} A., {Katz} N., {Wadsley} J., {Governato} F., {Quinn} T.,
  2006, MNRAS, 373, 1074

\bibitem[{{Sun} {et~al.}(2009){Sun}, {Voit}, {Donahue}, {Jones}, {Forman}, \&
  {Vikhlinin}}]{SVD09}
{Sun} M., {Voit} G.~M., {Donahue} M., {Jones} C., {Forman} W., {Vikhlinin} A.,
  2009, ApJ, 693, 1142

\bibitem[{{Sutherland} \& {Dopita}(1993)}]{SUD93}
{Sutherland} R.~S., {Dopita} M.~A., 1993, ApJ Supp., 88, 253

\bibitem[{{Tamura} {et~al.}(2001){Tamura}, {Bleeker}, {Kaastra}, {Ferrigno}, \&
  {Molendi}}]{TBK01}
{Tamura} T., {Bleeker} J.~A.~M., {Kaastra} J.~S., {Ferrigno} C., {Molendi} S.,
  2001, A\&A, 379, 107

\bibitem[{{Tamura} {et~al.}(2004){Tamura}, {Kaastra}, {den Herder}, {Bleeker},
  \& {Peterson}}]{TKH04}
{Tamura} T., {Kaastra} J.~S., {den Herder} J.~W.~A., {Bleeker} J.~A.~M.,
  {Peterson} J.~R., 2004, A\&A, 420, 135

\bibitem[{{Thacker} \& {Couchman}(2000)}]{THC00}
{Thacker} R.~J., {Couchman} H.~M.~P., 2000, ApJ, 545, 728

\bibitem[{{Tornatore} {et~al.}(2007){Tornatore}, {Borgani}, {Dolag}, \&
  {Matteucci}}]{TBD07}
{Tornatore} L., {Borgani} S., {Dolag} K., {Matteucci} F., 2007, MNRAS, 382,
  1050

\bibitem[{{Tornatore} {et~al.}(2004){Tornatore}, {Borgani}, {Matteucci},
  {Recchi}, \& {Tozzi}}]{TBM04}
{Tornatore} L., {Borgani} S., {Matteucci} F., {Recchi} S., {Tozzi} P., 2004,
  MNRAS, 349, L19

\bibitem[{{Tornatore} {et~al.}(2003){Tornatore}, {Borgani}, {Springel},
  {Matteucci}, {Menci}, \& {Murante}}]{TBS03}
{Tornatore} L., {Borgani} S., {Springel} V., {Matteucci} F., {Menci} N.,
  {Murante} G., 2003, MNRAS, 342, 1025

\bibitem[{{Tozzi} \& {Norman}(2001)}]{TON01}
{Tozzi} P., {Norman} C., 2001, ApJ, 546, 63

\bibitem[{Vacca {et~al.}(2012)Vacca, Murgia, Govoni, Feretti, Giovannini,
  Perley, \& Taylor}]{VMG12}
Vacca V., Murgia M., Govoni F., Feretti L., Giovannini G., Perley R.~A., Taylor
  G.~B., 2012, A\&A

\bibitem[{{Valdarnini}(2003)}]{VAL03}
{Valdarnini} R., 2003, MNRAS, 339, 1117

\bibitem[{Vazza {et~al.}(2012)Vazza, Bruggen, Gheller, \& Brunetti}]{VBG12}
Vazza F., Bruggen M., Gheller C., Brunetti G., 2012, MNRAS, 421, 3375

\bibitem[{{Veilleux} {et~al.}(2005){Veilleux}, {Cecil}, \&
  {Bland-Hawthorn}}]{VCB05}
{Veilleux} S., {Cecil} G., {Bland-Hawthorn} J., 2005, ARA\&A, 43, 769

\bibitem[{{Vikhlinin} {et~al.}(2006){Vikhlinin}, {Kravtsov}, {Forman}, {Jones},
  {Markevitch}, {Murray}, \& {Van Speybroeck}}]{VKF06}
{Vikhlinin} A., {Kravtsov} A., {Forman} W., {Jones} C., {Markevitch} M.,
  {Murray} S.~S., {Van Speybroeck} L., 2006, ApJ, 640, 691

\bibitem[{{Vikhlinin} {et~al.}(2005){Vikhlinin}, {Markevitch}, {Murray},
  {Jones}, {Forman}, \& {Van Speybroeck}}]{VMM05}
{Vikhlinin} A., {Markevitch} M., {Murray} S.~S., {Jones} C., {Forman} W., {Van
  Speybroeck} L., 2005, ApJ, 628, 655

\bibitem[{{Voit} {et~al.}(2005){Voit}, {Kay}, \& {Bryan}}]{VKB05}
{Voit} G.~M., {Kay} S.~T., {Bryan} G.~L., 2005, MNRAS, 364, 909

\bibitem[{{Wiersma} {et~al.}(2011){Wiersma}, {Schaye}, \& {Theuns}}]{WST11}
{Wiersma} R.~P.~C., {Schaye} J., {Theuns} T., 2011, MNRAS, 415, 353

\bibitem[{{Young} {et~al.}(2011){Young}, {Thomas}, {Short}, \&
  {Pearce}}]{YTS11}
{Young} O.~E., {Thomas} P.~A., {Short} C.~J., {Pearce} F., 2011, MNRAS, 413,
  691

\end{thebibliography}

\end{document}